\documentclass[prd,twocolumn,showpacs,floatfix,amsmath,amssymb,floatfix,nofootinbib]{revtex4}
\usepackage{graphicx,color,dcolumn,booktabs,bm}
\usepackage{longtable,lscape,comment,braket}
\usepackage{txfonts}
\usepackage{overpic}
\usepackage{amssymb}
\usepackage{indentfirst}
\usepackage{feynmf}   
\usepackage{slashed}  
\usepackage{cases}
\usepackage{epstopdf}
\usepackage{psfrag}
\usepackage{subfigure}
\usepackage{threeparttable}
\pdfoutput=1
\usepackage{color}
\usepackage{multirow}

\def\lrpartial{\buildrel\leftrightarrow\over\partial}
\graphicspath{{Figures/}} %
\usepackage[colorlinks, citecolor=blue,anchorcolor=red,menucolor=red, linkcolor=red,filecolor=red,runcolor=red,urlcolor=blue,frenchlinks=red]{hyperref}

\usepackage{ulem}
\def\be{\begin{equation}}
\def\ee{\end{equation}}

\begin{document}

\title{Exploring the $\Upsilon(6S)\to \chi_{bJ}\phi$ and $\Upsilon(6S)\to \chi_{bJ}\omega$ hidden-bottom hadronic transitions}
\author{Qi Huang}\email{huangq16@lzu.edu.cn}
\author{Bo Wang}\email{wangb13@lzu.edu.cn}
\author{Xiang Liu}\email{xiangliu@lzu.edu.cn}
\affiliation{School of Physical Science and Technology, Lanzhou University, Lanzhou 730000, China\\
Research Center for Hadron and CSR Physics, Lanzhou University and Institute of Modern Physics of CAS, Lanzhou 730000, China}

\author{Dian-Yong Chen}\email{chendy@seu.edu.cn}
\affiliation{Department of Physics, Southeast University, Nanjing 210094, China}

\author{Takayuki Matsuki}\email{matsuki@tokyo-kasei.ac.jp}
\affiliation{Tokyo Kasei University, 1-18-1 Kaga, Itabashi, Tokyo 173-8602, Japan\\
Theoretical Research Division, Nishina Center, RIKEN, Wako, Saitama 351-0198, Japan}
\begin{abstract}
In this work, we investigate the hadronic loop contributions to the $\Upsilon(6S) \to \chi_{bJ} \phi~(J=0,1,2)$ along with $\Upsilon(6S) \to \chi_{bJ} \omega~(J=0,1,2)$ transitions. We predict that the branching ratios of $\Upsilon(6S) \to \chi_{b0} \phi$, $\Upsilon(6S) \to \chi_{b1} \phi$ and $\Upsilon(6S) \to \chi_{b2} \phi$ are $(0.68 \sim 4.62) \times 10^{-6}$, $(0.50 \sim 3.43) \times 10^{-6}$ and $(2.22 \sim 15.18) \times 10^{-6}$, respectively and those of $\Upsilon(6S) \to \chi_{b0} \omega$, $\Upsilon(6S) \to \chi_{b1} \omega$ and $\Upsilon(6S) \to \chi_{b2} \omega$ are $(0.15 \sim 2.81) \times 10^{-3}$, $(0.63 \sim 11.68) \times 10^{-3}$ and $(1.08 \sim 20.02) \times 10^{-3}$, respectively. Especially, some typical ratios, which reflect the relative magnitudes of the predicted branching ratios, are given, i.e., for $\Upsilon(6S)\to \chi_{bJ}\phi$ transitions, 
$\mathcal{R}^\phi_{10}={\mathcal{B}[\Upsilon(6S) \to \chi_{b1} \phi]}/{\mathcal{B}[\Upsilon(6S) \to \chi_{b0} \phi]} \approx 0.74$, $\mathcal{R}^\phi_{20}= {\mathcal{B}[\Upsilon(6S) \to \chi_{b2} \phi]}/{\mathcal{B}[\Upsilon(6S) \to \chi_{b0} \phi]} \approx 3.28$, and $\mathcal{R}^\phi_{21} = {\mathcal{B}[\Upsilon(6S) \to \chi_{b2} \phi]}/{\mathcal{B}[\Upsilon(6S) \to \chi_{b1} \phi]} \approx 4.43$, and for $\Upsilon(6S)\to \chi_{bJ}\omega$ transitions, $\mathcal{R}^\omega_{10}={\mathcal{B}[\Upsilon(6S) \to \chi_{b1} \omega]}/{\mathcal{B}[\Upsilon(6S) \to \chi_{b0} \omega]} \approx 4.11$, $\mathcal{R}^\omega_{20}= {\mathcal{B}[\Upsilon(6S) \to \chi_{b2} \omega]}/{\mathcal{B}[\Upsilon(6S) \to \chi_{b0} \omega]} \approx 7.06$, and $\mathcal{R}^\omega_{21} = {\mathcal{B}[\Upsilon(6S) \to \chi_{b2} \omega]}/{\mathcal{B}[\Upsilon(6S) \to \chi_{b1} \omega]} \approx 1.72$. With the running of BelleII in the near future, experimental measurement of these two kinds of transitions will be a potential research issue.
\end{abstract}

 \maketitle

\section{Introduction}\label{sec1}

As an interesting research issue, experimental studies of the hadronic transitions of $\Upsilon(5S)$ have been focused on by the Belle Collaboration in the past decade. When surveying the reported hadronic transitions of $\Upsilon(5S)$, we found their general property, i.e., their observed  hadronic transitions have large branching ratios. For example, Belle observed anomalous decay widths of the $\Upsilon(5S) \to \Upsilon(nS) \pi^+ \pi^-$ \cite{Abe:2007tk}, and $\Upsilon(5S) \to \chi_{bJ} \omega~(J=0,1,2)$ transitions \cite{He:2014sqj}. In addition, two bottomonium-like states $Z_b(10610)$ and $Z_b(10650)$ were observed in $\Upsilon(5S) \to \Upsilon(nS) \pi^+ \pi^-$ \cite{Belle:2011aa}.
As indicated in a serial of theoretical studies \cite{Chen:2011zv,Meng:2007tk,Meng:2008dd,Simonov:2008qy,Chen:2011qx,Chen:2014ccr,Chen:2011pv}, the puzzling phenomena happening on $\Upsilon(5S)$ transitions reflect an underlying mechanism mediated by a coupled channel effect since $\Upsilon(5S)$ is above the thresholds of $B_{(s)}^{(*)}\bar{B}_{(s)}^{(*)}$ \cite{Olive:2016xmw}.

In the bottomonium family, the $\Upsilon(6S)$ has the similar situation to that of $\Upsilon(5S)$. We have a reason to believe that the coupled channel effect is still important to the hadronic transitions of $\Upsilon(6S)$, whose exploration is, thus, an intriguing topic. This theme can provide us a valuable information of the coupled-channel effect on these decays.

In this work, we calculate the $\Upsilon(6S)\to \chi_{bJ}\phi$ ($J=0,1,2$) along with $\Upsilon(6S)\to \chi_{bJ}\omega$ ($J=0,1,2$) processes via the hadronic loop mechanism, which is an equivalent description of the coupled channel effect \cite{Chen:2011zv,Meng:2007tk,Meng:2008dd,Chen:2011qx,Chen:2014ccr,Chen:2011pv,Meng:2008bq,Wang:2016qmz,Liu:2006dq, Liu:2009dr,Li:2013zcr}. By analyzing these transitions, the relative decay rates of $\Upsilon(6S)\to \chi_{bJ}\phi$ ($J=0,1,2$) and $\Upsilon(6S)\to \chi_{bJ}\omega$ ($J=0,1,2$), which are a typical physical quantity given by our calculation, are determined. Especially, our results show that these relative decay rates are weakly dependent on the model parameters.
Thus, experimental measurement of these rates can be a crucial test of the hadronic loop mechanism in the $\Upsilon(6S)\to \chi_{bJ}\phi$ and $\Upsilon(6S)\to \chi_{bJ}\omega$ decays. In addition, we also estimate the typical values of the branching ratios of $\Upsilon(6S)\to \chi_{bJ}\phi$ and $\Upsilon(6S)\to \chi_{bJ}\omega$, which can be measured experimentally in near future. Anyway,  we would like to inspire experimenlists' interest in searching for the $\Upsilon(6S)\to \chi_{bJ}\phi$ and $\Upsilon(6S)\to \chi_{bJ}\omega$ decays by our results presented in this work.

This paper is organized as follows. After introduction, we present the detailed calculation of $\Upsilon(6S)\to\chi_{bJ}\phi$ and $\Upsilon(6S)\to\chi_{bJ}\omega$ via the hadronic loop mechanism in Sec. \ref{sec2}. The numerical results are presented in Sec. \ref{sec3}. The paper ends with a short summary.

\section{${\Upsilon(6S) \to  \chi_{bJ}}\phi$ and ${\Upsilon(6S) \to  \chi_{bJ}}\omega$ transitions via Hadronic Loop Mechanism}\label{sec2}

\begin{center}
\begin{figure}[htbp]
\scalebox{0.075}{\includegraphics{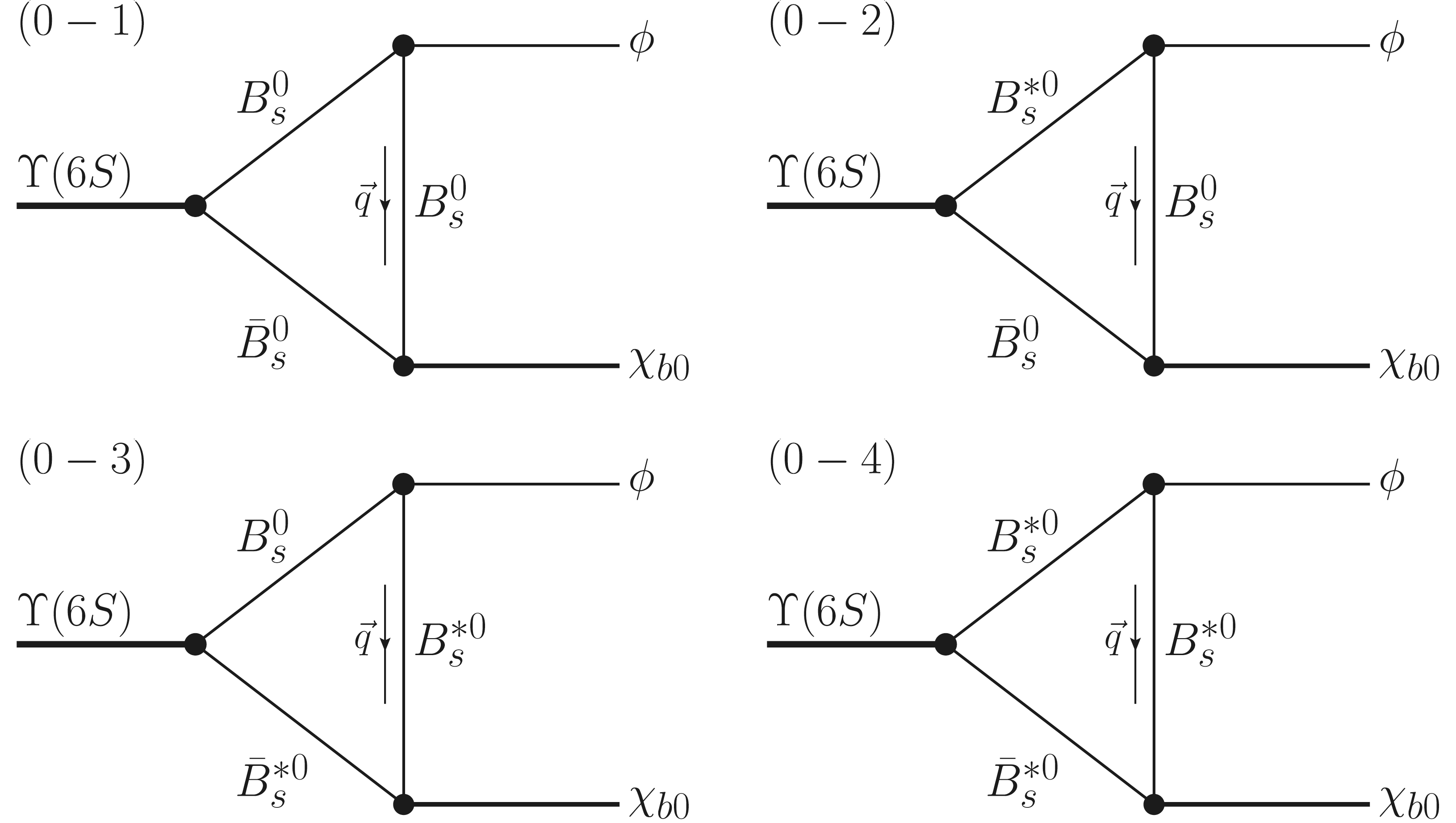}}
\caption{Schematic diagrams depicting the ${\Upsilon(6S) \to  \chi_{b0}\phi}$ process via the hadronic loop mechanism.}
\label{fig:6S-phi-chib0}
\end{figure}
\end{center}

\begin{center}
\begin{figure}[htbp]
\scalebox{0.075}{\includegraphics{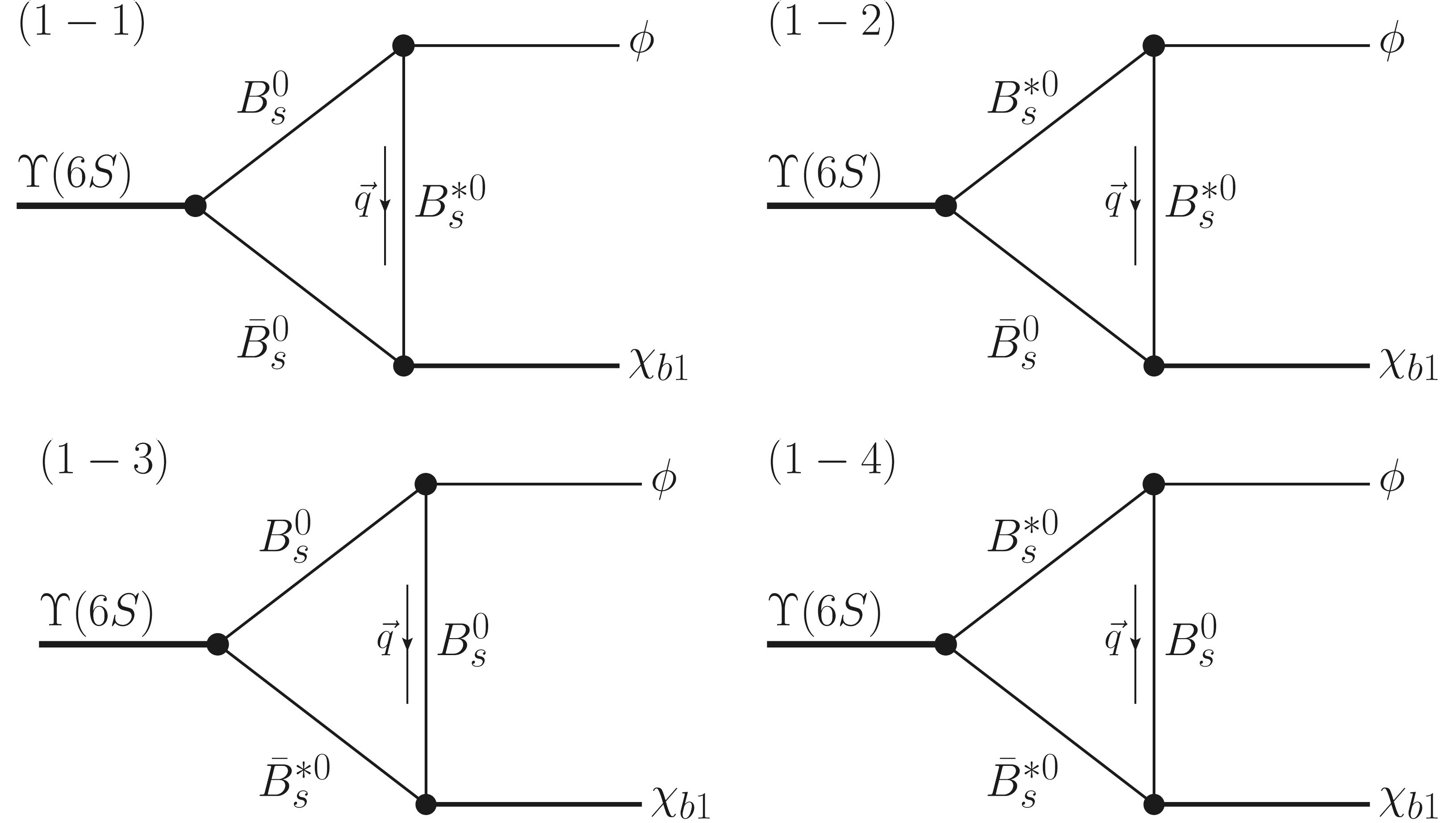}}
\caption{Schematic diagrams depicting the ${\Upsilon(6S) \to  \chi_{b1}\phi}$ process via the hadronic loop mechanism.}
\label{fig:6S-phi-chib1}
\end{figure}
\end{center}

\begin{center}
\begin{figure}[htbp]
\scalebox{0.075}{\includegraphics{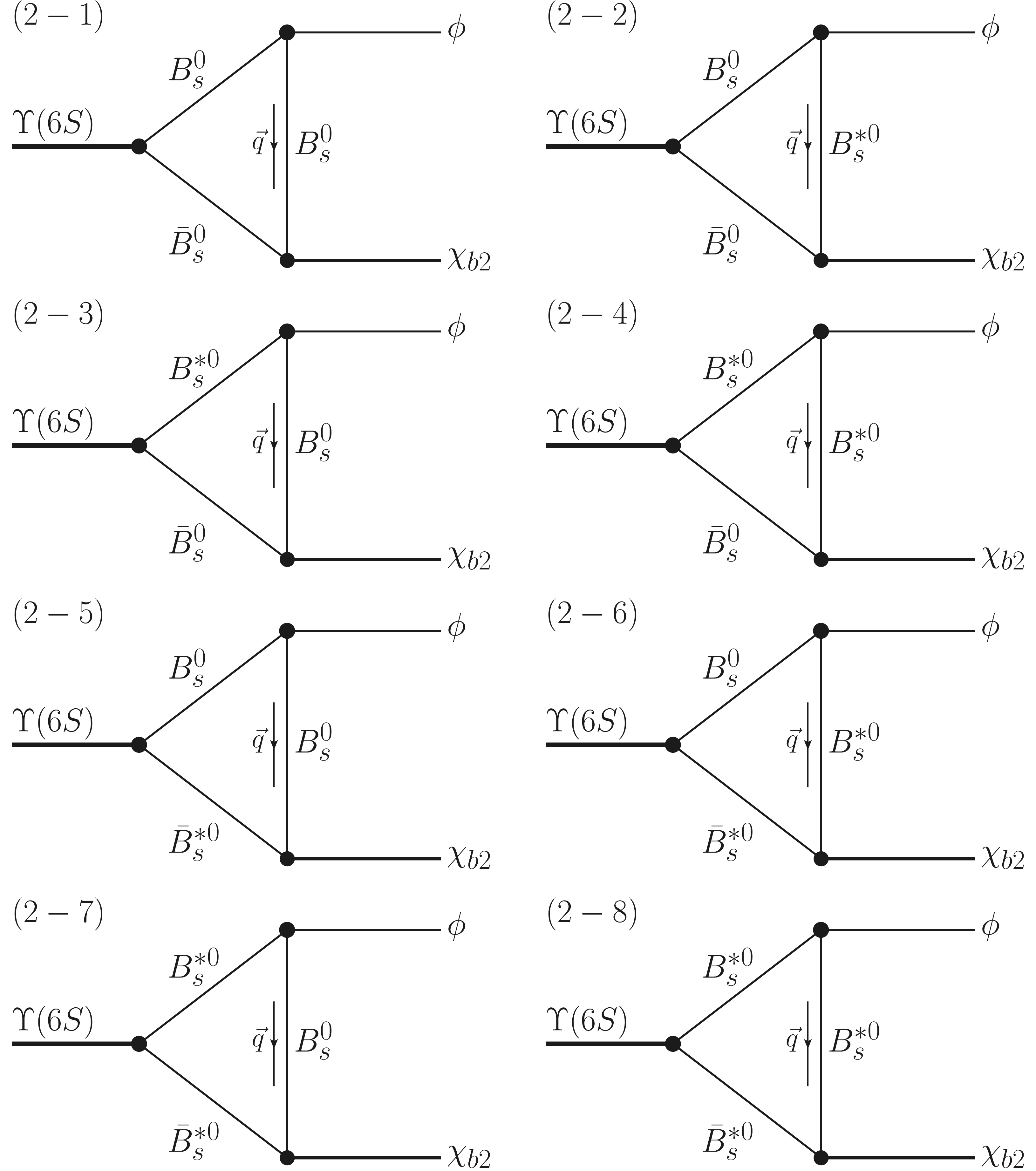}}
\caption{Schematic diagrams depicting the ${\Upsilon(6S) \to  \chi_{b2}\phi}$ process via the hadronic loop mechanism.}
\label{fig:6S-phi-chib2}
\end{figure}
\end{center}

\begin{center}
\begin{figure}[htbp]
\scalebox{0.075}{\includegraphics{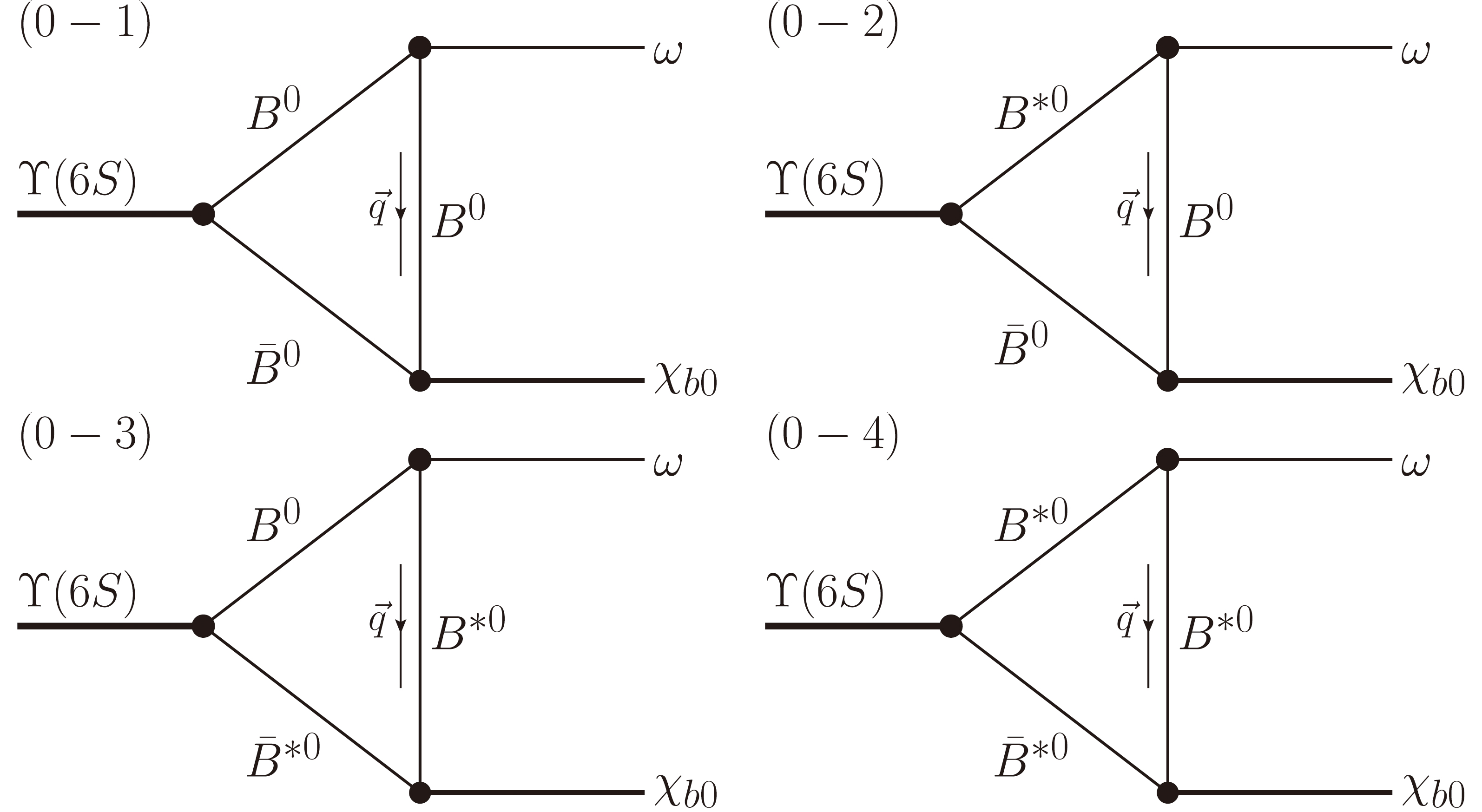}}
\caption{Schematic diagrams depicting the ${\Upsilon(6S) \to  \chi_{b0}\omega}$ process via the hadronic loop mechanism.}
\label{fig:6S-omega-chib0}
\end{figure}
\end{center}

\begin{center}
\begin{figure}[htbp]
\scalebox{0.075}{\includegraphics{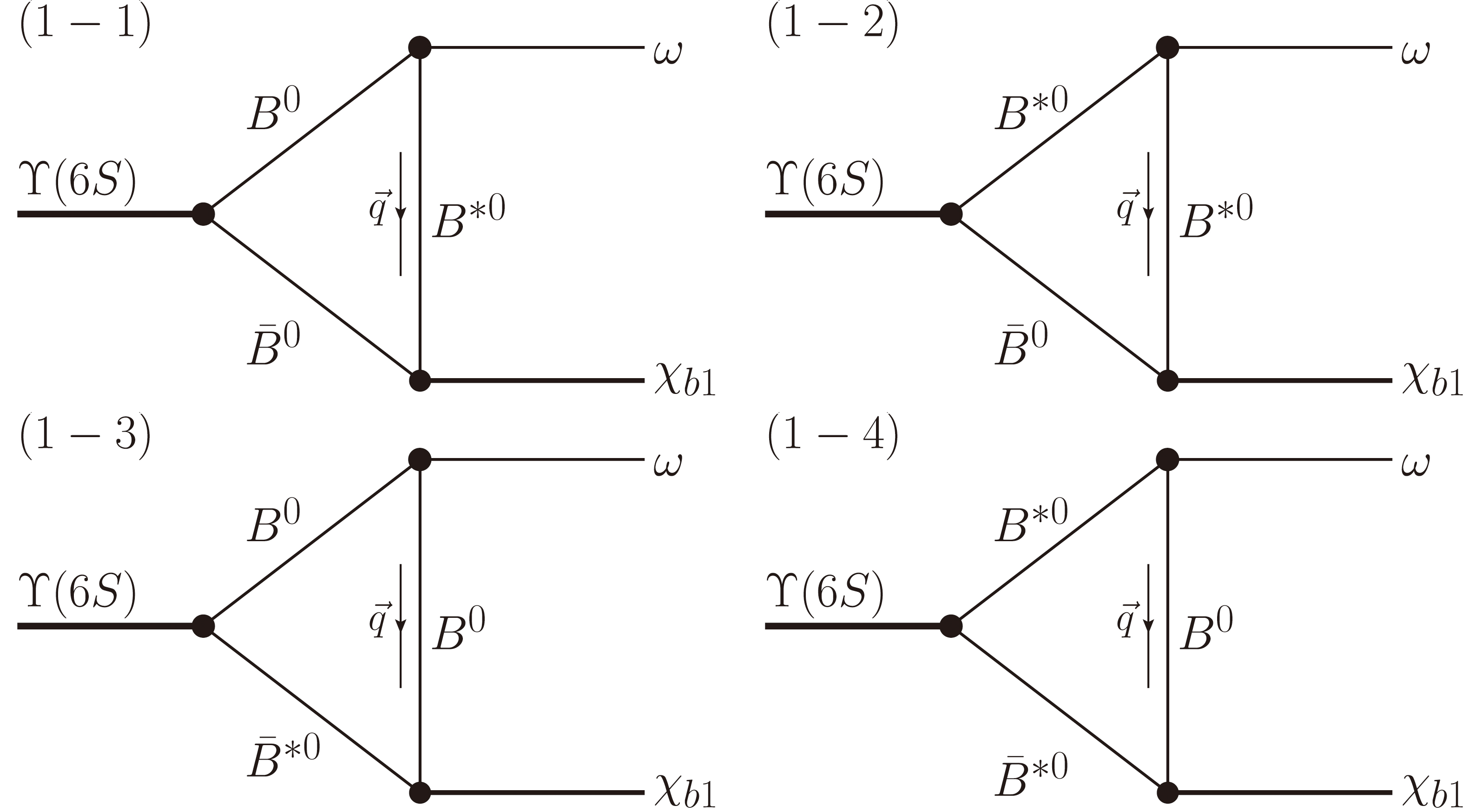}}
\caption{Schematic diagrams depicting the ${\Upsilon(6S) \to  \chi_{b1}\omega}$ process via the hadronic loop mechanism.}
\label{fig:6S-omega-chib1}
\end{figure}
\end{center}

\begin{center}
\begin{figure}[htbp]
\scalebox{0.075}{\includegraphics{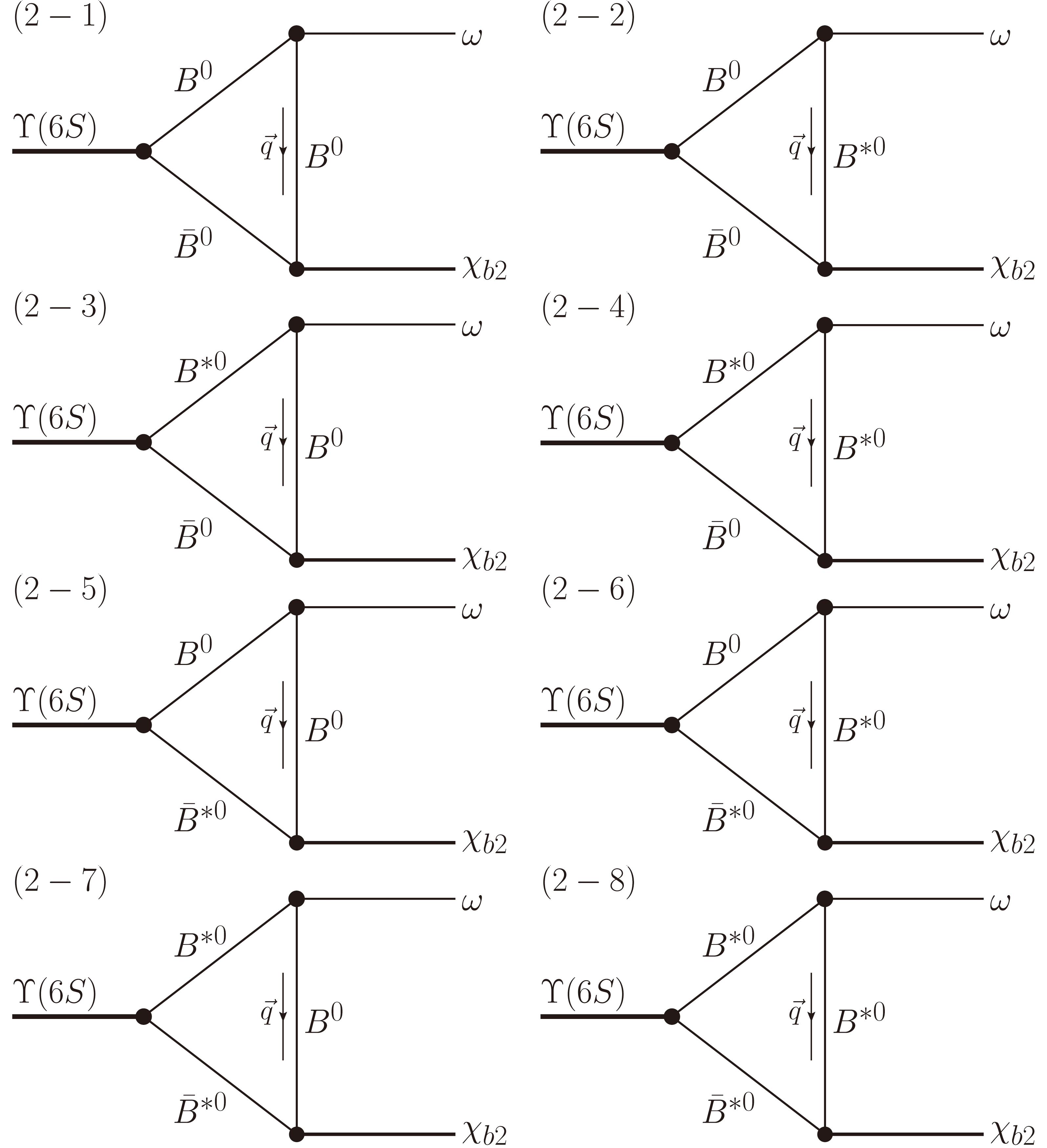}}
\caption{Schematic diagrams depicting the ${\Upsilon(6S) \to  \chi_{b2}\omega}$ process via the hadronic loop mechanism.}
\label{fig:6S-omega-chib2}
\end{figure}
\end{center}

Under the hadronic loop mechanism, the ${\Upsilon(6S) \to  \chi_{bJ}}\phi$ transitions occur via the triangle loops composed of $B_s^{(*)0}$ and $\bar B_{s}^{(*)0}$, which play a role of the bridge to connect the initial state $\Upsilon(6S)$ and final states $\phi$ and $\chi_{bJ}$. In Figs. \ref{fig:6S-phi-chib0}-\ref{fig:6S-phi-chib2}, we list the typical diagrams depicting the $\Upsilon(6S) \to \chi_{bJ} \phi\,(J=0,1,2)$ transitions. For the ${\Upsilon(6S) \to  \chi_{bJ}}\omega$ transitions, due to very different quark contents between $\phi$ and $\omega$, the bridges change to $B^{(*)}$ and $\bar{B}^{(*)}$ and the diagrams change simultaneously as in Figs. \ref{fig:6S-omega-chib0}-\ref{fig:6S-omega-chib2}.

To calculate these diagrams at the hadron level, we adopt the effective Lagrangian approach, in which we first introduce the Lagrangians relevant to our calculation.

For the interactions between a heavy quarkonium and two heavy-light mesons, the Lagrangians are constructed based on the heavy quark effective theory. In the heavy quark limit, the light degrees of freedom $s_\ell$ is a good quantum number. Thus, each value of $s_\ell$ is assigned to a doublet formed by the states with a total angular momentum $J=s_\ell \pm 1/2$,
while for the heavy quarkonium, since the degeneracy is expected under the rotations of two heavy quark spins, there is a multiplet formed by heavy quarkonia with the same angular momentum $\ell$.

Therefore, under the framework of heavy quark symmetry, general forms of couplings between an S-wave or P-wave heavy quarkonium and two heavy-light mesons can be constructed as \cite{Casalbuoni:1996pg}
\begin{eqnarray}\label{eqs:SPHH}
\mathcal{L}_s&=& ig \mathrm{Tr} \left[R^{(Q\bar{Q})} \bar{H}^{(\bar{Q}q)} \gamma^\mu \lrpartial_\mu \bar{H}^{(Q\bar{q})} \right] +H.c., \nonumber\\
\mathcal{L}_p &=& ig_1 \mathrm{Tr} \left[P^{(Q\bar{Q}) \mu} \bar{H}^{(\bar{Q}q)} \gamma_\mu \bar{H}^{(Q\bar{q})}\right] +H.c.,
\end{eqnarray}
in which ${R^{(Q\bar{Q})}}$ and ${P^{(Q\bar{Q})}}$ denote multiplets formed by bottomonia with ${\ell=0}$ and ${\ell=1}$, and their detailed expressions, as in Ref. \cite{Colangelo:2002mj}, can be written as
\begin{eqnarray}
R^{(Q\bar{Q})} &=& \frac{1+\slashed{v}}{2} \left[\Upsilon^\mu \gamma_\mu -\eta_b \gamma_5\right]\frac{1-\slashed{v}}{2},
\end{eqnarray}
\begin{eqnarray}
{P^{(Q\bar{Q})}}^\mu &=& \frac{1+\slashed{v}}{2} \Big[\chi_{b2}^{\mu \alpha} \gamma_\alpha +\frac{1}{\sqrt{2}}  \varepsilon^{\mu \alpha \beta \gamma} v_\alpha \gamma_\beta \chi_{b1 \gamma} \nonumber\\
&&+\frac{1}{\sqrt{3}} \big(\gamma^\mu -v^\mu \big) \chi_{b0} +h_b^\mu \gamma_5 \Big] \frac{1-\slashed{v}}{2},
\end{eqnarray}
respectively. ${H^{(Q\bar{q})}}$ represents a doublet formed by heavy-light pseudoscalar and vector mesons \cite{Casalbuoni:1996pg,Kaymakcalan:1983qq,Oh:2000qr,Colangelo:2002mj}
\begin{eqnarray}
H^{(Q\bar{q})}=\frac{1+ \slashed{v}}{2} \left[\mathcal{B}^\ast_\mu \gamma^\mu-\mathcal{B} \gamma^5\right],
\end{eqnarray}
with definitions $\mathcal{B}^{(*)\dag} = (B^{(*)+},B^{(*)0},B_s^{(*)0})$ and $\mathcal{B}^{(*)} = (B^{(*)-},\bar{B}^{(*)0},\bar{B}_s^{(*)0})^T$ as in Ref. \cite{Wang:2016qmz}. $H^{(\bar{Q}q)}$ corresponds to a doublet formed by 
heavy-light anti-mesons, which can be obtained by applying the charge conjugation operation to $H^{(Q\bar{q})}$.

For the interaction between a light vector meson and two heavy-light mesons, the general form of the Lagrangian reads as \cite{Casalbuoni:1996pg,Cheng:1992xi,Yan:1992gz,Wise:1992hn,Burdman:1992gh,Falk}
\begin{eqnarray}\label{eqs:HHV}
\mathcal{L}_\mathcal{V} = i\beta \mathrm{Tr}[H^j v^\mu (-\rho_\mu)_j^i \bar{H}_i] + i\lambda \mathrm{Tr}[H^j \sigma^{\mu\nu} F_{\mu\nu} (\rho) \bar{H}_i],
\end{eqnarray}
where
\begin{eqnarray}
\rho_\mu &=& i \frac{g_V}{\sqrt{2}} \mathcal{V}_\mu,\\
F_{\mu\nu} (\rho) &=& \partial_\mu \rho_\nu - \partial_\nu \rho_\mu + [\rho_\mu,\rho_\nu],
\end{eqnarray}
and a vector octet $\mathcal{V}$ has the form
\begin{eqnarray}
\mathcal{V} &=&
 \left(
 \begin{array}{ccc}
\frac{1}{\sqrt{2}} (\rho^{0}+ \omega) & \rho^{+} & K^{*+}\\
\rho^{-} & \frac{1}{\sqrt{2}}(-\rho^0 +\omega) &  K^{*0}\\
 K^{*-} & \bar{K}^{*0} & \phi
 \end{array}
 \right).
\end{eqnarray}

By expanding the Lagrangians in Eqs. (\ref{eqs:SPHH}) and (\ref{eqs:HHV}), the following concrete expressions are obtained
\begin{eqnarray}
&&\mathcal{L}_{\Upsilon \mathcal{B}^{(\ast)} \mathcal{B}^{(\ast)}}\nonumber\\&&
= -ig_{\Upsilon \mathcal{BB} } \Upsilon_\mu (\partial^\mu
\mathcal{B} \mathcal{B}^\dagger- \mathcal{B}
\partial^\mu \mathcal{B}^\dagger) \nonumber\\ && \quad+ g_{\Upsilon
\mathcal{B}^\ast \mathcal{B}} \varepsilon^{\mu \nu \alpha \beta}
\partial_\mu \Upsilon_\nu (\mathcal{B}^\ast_\alpha \lrpartial_\beta
\mathcal{B}^\dagger -\mathcal{B} \lrpartial_\beta
\mathcal{B}_\alpha^{\ast \dagger} ) \nonumber\\ && \quad+ ig_{\Upsilon
\mathcal{B}^\ast \mathcal{B}^\ast} \Upsilon^\mu
(\mathcal{B}^\ast_\nu \partial^\nu \mathcal{B}^{\ast \dagger}_\mu
-\partial^\nu \mathcal{B}^{\ast}_\mu \mathcal{B}^{\ast \dagger}_\nu
-\mathcal{B}^\ast_\nu \lrpartial_\mu \mathcal{B}^{\ast \nu
\dagger}),
\end{eqnarray}
\begin{eqnarray}
&&\mathcal{L}_{\chi_{bJ} \mathcal{B}^{(\ast)} \mathcal{B}^{(\ast)}}\nonumber\\
&&= - g_{\chi_{b0} \mathcal{B} \mathcal{B} } \chi_{b0} \mathcal{B}
\mathcal{B}^\dagger - g_{\chi_{b0} \mathcal{B}^\ast
\mathcal{B}^\ast} \chi_{b0} \mathcal{B}_{\mu}^\ast \mathcal{B}^{\ast
\mu\dagger } \nonumber\\ && \quad +i g_{\chi_{c1} \mathcal{B}
\mathcal{B}^\ast} \chi_{b1}^\mu ( \mathcal{B}^{\ast }_\mu
\mathcal{B}^\dagger - \mathcal{B} \mathcal{B}^{\ast \dagger}_\mu )
\nonumber\\
&&\quad- g_{\chi_{b2} \mathcal{B} \mathcal{B}}
\chi_{b2}^{\mu \nu}
\partial_\mu \mathcal{B} \partial_\nu \mathcal{B}^\dagger\nonumber\\
&&\quad+ g_{\chi_{b2} \mathcal{B}^\ast \mathcal{B}^\ast} \chi_{b2}^{\mu
\nu}
(\mathcal{B}^\ast_{\mu} \mathcal{B}^{\ast \dagger}_\nu + \mathcal{B}^\ast_{\nu} \mathcal{B}^{\ast \dagger}_\mu) \nonumber\\
&&  \quad-ig_{\chi_{b2} \mathcal{B}^\ast \mathcal{B}} \varepsilon_{\mu
\nu \alpha \beta} \partial^\alpha \chi_{b2}^{\mu \rho}
(\partial_\rho \mathcal{B}^{\ast \nu} \partial^\beta
\mathcal{B}^\dagger -\partial^\beta \mathcal{B}
\partial_\rho \mathcal{B}^{\ast \nu \dagger} ),
\end{eqnarray}
\begin{eqnarray}
&&\mathcal{L}_{\mathcal{B}^{(\ast)}\mathcal{B}^{(\ast)} \mathcal{V}}
\nonumber\\&&= -ig_{\mathcal{B} \mathcal{B}\mathcal{V}} \mathcal{B}_i^\dagger
\lrpartial^\mu \mathcal{B}^j (\mathcal{V}_\mu)^i_j\nonumber\\
&&\quad-2 f_{\mathcal{B}^\ast \mathcal{B} \mathcal{V}} \varepsilon_{\mu \nu
\alpha \beta} (\partial^\mu
\mathcal{V}^\nu)^i_j (\mathcal{B}^\dagger_i \lrpartial^\alpha
\mathcal{B}^{\ast \beta j} -\mathcal{B}_i^{\ast \beta \dagger}
\lrpartial^\alpha \mathcal{B}^j) \nonumber\\&& \quad+ig_{\mathcal{B}^\ast
\mathcal{B}^\ast \mathcal{V}} \mathcal{B}^{\ast \nu \dagger}_i
\lrpartial^\mu \mathcal{B}^{\ast j}_\nu (\mathcal{V}_\mu)^i_j
\nonumber\\ &&\quad+4if_{\mathcal{B}^\ast \mathcal{B}^\ast \mathcal{V}}
\mathcal{B}^{\ast \dagger}_{i\mu} (\partial^\mu \mathcal{V}^\nu
-\partial^\nu \mathcal{V}^\mu)^i_j \mathcal{B}^{\ast j}_\nu.
\end{eqnarray}

With the above effective Lagrangians, we can write out the amplitudes of hadronic loop contributions to $\Upsilon(6S) \to \chi_{bJ} \phi\,(J=0,1,2)$. For the $\Upsilon(6S) \to \chi_{b0} \phi$ transition, the amplitudes corresponding to Fig. \ref{fig:6S-phi-chib0} are
\begin{eqnarray}
\mathcal{M}_{(0-1)} &=& \int \frac{d^4q}{(2\pi)^4} [-i g_{\Upsilon B_s B_s} \epsilon_\Upsilon^\mu ((i k_1)_\mu - (i k_2)_\mu)]\nonumber\\
&&\times [-i g_{B_s B_s \phi} \epsilon^*_{\phi\lambda} ((-i k_1)^\lambda -(i q)^\lambda)] [-g_{B_s B_s \chi_{b0}}]\nonumber\\
&&\times \frac{1}{k_1^2-m_{B_s}^2} \frac{1}{k_2^2-m_{B_s}^2} \frac{1}{q^2-m_{B_s}^2}\mathcal{F}^2(q^2),
\end{eqnarray}

\begin{eqnarray}
\mathcal{M}_{(0-2)} &=& \int \frac{d^4q}{(2\pi)^4} [-g_{\Upsilon B_s^* B_s} \varepsilon^{\mu\nu\alpha\beta} \epsilon_{\Upsilon\mu} (-i p_1)_\nu ((i k_2)_\beta - (i k_1)_\beta)]\nonumber\\
&&\times [2 f_{B_s^* B_s \phi} \varepsilon_{\lambda\rho\delta\sigma} \epsilon_\phi^{*\lambda} (i p_2)^\rho ((-i k_1)^\delta - (i q)^\delta)] [-g_{B_s B_s \chi_{b0}}]\nonumber\\
&&\times \frac{-g_\alpha^\sigma + k_{1\alpha} k_1^\sigma / m_{B_s^*}^2}{k_1^2-m_{B_s^*}^2} \frac{1}{k_2^2-m_{B_s}^2} \frac{1}{q^2-m_{B_s}^2}\mathcal{F}^2(q^2),
\end{eqnarray}

\begin{eqnarray}
\mathcal{M}_{(0-3)} &=& \int \frac{d^4q}{(2\pi)^4} [g_{\Upsilon B_s B_s^*} \varepsilon^{\mu\nu\alpha\beta} \epsilon_{\Upsilon\mu} (-i p_1)_\nu ((i k_2)_\beta - (i k_1)_\beta)]\nonumber\\
&&\times [-2 f_{B_s B_s^* \phi} \varepsilon_{\lambda\rho\delta\sigma} \epsilon_\phi^{*\lambda} (i p_2)^\rho ((-i k_1)^\delta - (i q)^\delta)]\nonumber\\
&&\times [-g_{B_s^* B_s^* \chi_{b0}}]\frac{1}{k_1^2-m_{B_s}^2} \frac{-g_\alpha^\zeta + k_{2\alpha} k_2^\zeta / m_{B_s^*}^2}{k_2^2-m_{B_s^*}^2}\nonumber\\
&&\times \frac{-g^\sigma_\zeta + q^\sigma q_\zeta / m_{B_s^*}^2 }{q^2-m_{B_s^*}^2}\mathcal{F}^2(q^2),
\end{eqnarray}

\begin{eqnarray}
\mathcal{M}_{(0-4)} &=& \int \frac{d^4q}{(2\pi)^4} [i g_{\Upsilon B_s^* B_s^*} \epsilon_\Upsilon^\mu (g_{\nu\alpha} g_{\mu\beta} (i k_2)^\nu - g_{\mu\alpha} g_{\nu\beta} (i k_1)^\nu\nonumber\\
&& - g_{\alpha\beta}((i k_2)_\mu - (i k_1)_\mu))]\nonumber\\
&&\times [\epsilon^*_{\phi\lambda} (i g_{B_s^* B_s^* \phi} g^{\delta\sigma} ((-i k_1)^\lambda -(i q)^\lambda)\nonumber\\
&&+ 4 i f_{B_s^* B_s^* \phi} (i p_2)_\rho (g^{\lambda\delta} g^{\rho\sigma} - g^{\lambda\sigma} g^{\rho\delta}))] [-g_{B_s^* B_s^* \chi_{b0}}]\nonumber\\
&&\times \frac{-g_\delta^\alpha + k_{1\delta} k_1^\alpha / m_{B_s^*}^2}{k_1^2-m_{B_s^*}^2} \frac{-g^{\beta\zeta} + k_2^\beta k_2^\zeta / m_{B_s^*}^2}{k_2^2-m_{B_s^*}^2}\nonumber\\
&&\times \frac{-g_{\sigma\zeta} + q_\sigma q_\zeta / m_{B_s^*}^2}{q^2-m_{B_s^*}^2}\mathcal{F}^2(q^2),
\end{eqnarray}
where $p_1$, $p_2$ and $p_3$ are momenta of $\Upsilon(6S)$, $\phi/\omega$ and $\chi_{bJ}$, and $k_1$, $k_2$, and $q$ are momenta of internal $B^{(*)}_{(s)}$ and exchanged $B^{(*)}_{(s)}$, respectively.
In these expressions of the decay amplitudes, the monopole form factor is introduced, by which the inner structure of interaction vertices is reflected and the off-shell effect of the exchanged bottom-strange mesons is compensated.
Here, the adopted form factor is taken as $\mathcal{F}(q^2) = (m_E^2 - \Lambda^2)/(q^2 - \Lambda^2)$, with $m_E$ being the mass of the exchanged boson and the cutoff $\Lambda$ being parameterized as $\Lambda = m_E + \alpha_\Lambda \Lambda_{QCD}$ with $\Lambda_{QCD}=0.22$ GeV as in Refs. \cite{Liu:2006dq,Liu:2009dr,Li:2013zcr}. {We need to specify that the monopole behavior of the adopted form factor
was suggested by the QCD sum rule studied in Ref. \cite{Gortchakov:1995im}. In a serial of published papers (see Refs. \cite{Chen:2011zv,Meng:2007tk,Meng:2008dd, Chen:2011qx,Chen:2014ccr,Chen:2011pv,Meng:2008bq,Wang:2016qmz,Liu:2006dq,Liu:2009dr,Li:2013zcr,Colangelo:2002mj,Cheng:2004ru}), the monopole form factor was adopted to study the transitions of charmonia and bottomonia, and $B$ decays.  Thus, this approach has been tested by these successful studies. }

In a similar way, we can further write out the decay amplitudes of $\Upsilon(6S) \to \chi_{b1} \phi$ and $\Upsilon(6S) \to \chi_{b2} \phi$, which are collected in Appendix.
By considering the isospin symmetry, a general expression of the total amplitude of ${\Upsilon(6S) \to \chi_{bJ} \phi}$ with ${J=0,1,2}$ is written as
\begin{eqnarray}
\mathcal{M}^{\mathrm{Total}}_{\Upsilon(6S)\to \chi_{bJ} \phi} = 2 \sum_{j} \mathcal{M}_{(J-j)} .
\end{eqnarray}
 Then, the partial decay width reads
\begin{eqnarray}
\Gamma_{\Upsilon(6S) \to \chi_{bJ} \phi} = \frac{1}{3} \frac{1}{8\pi} \frac{|\vec{p}_\phi|}{m_{\Upsilon(6S)}^2}
|\overline{\mathcal{M}^{\mathrm{Total}}_{\Upsilon(6S) \to \chi_{bJ} \phi}}|^2,
\end{eqnarray}
where the overline indicates the sum over polarizations of $\Upsilon(6S)$, $\phi$, and $\chi_{b1}$ (or $\chi_{b2}$) and the factor $\frac{1}{3}$ denotes the average over the polarization of initial $\Upsilon(6S)$.

In the case of $\Upsilon(6S) \to \chi_{bJ} \omega$, the expression of the partial decay width is given by
\begin{eqnarray}
\Gamma_{\Upsilon(6S) \to \chi_{bJ} \omega} = \frac{1}{3} \frac{1}{8\pi} \frac{|\vec{p}_\omega|}{m_{\Upsilon(6S)}^2}
|\overline{\mathcal{A}^{\mathrm{Total}}_{\Upsilon(6S) \to \chi_{bJ} \omega}}|^2,
\end{eqnarray}
with a general expression of the total amplitude of ${\Upsilon(6S) \to \chi_{bJ} \omega}$ as
\begin{eqnarray}
\mathcal{A}^{\mathrm{Total}}_{\Upsilon(6S)\to \chi_{bJ} \omega} = 4 \sum_{j} \mathcal{A}_{(J-j)}
\end{eqnarray}
by considering the isospin and charge symmetry. The detailed expressions of $\mathcal{A}_{(J-j)}$ are collected in Appendix.

\section{Numerical Results}\label{sec3}

With the formulas listed in Sec. \ref{sec2} and Appendix, we estimate the hadronic loop contributions to the $\Upsilon(6S) \to \chi_{bJ} \phi$ together with $\Upsilon(6S) \to \chi_{bJ} \omega~(J=0,1,2)$ transitions. Besides the masses taken from the Particle Data Book \cite{Olive:2016xmw}, all the other input parameters we need are the coupling constants. Since the $\Upsilon(6S)$ is above the threshold of $B_{(s)}^{(*)}\bar{B}_{(s)}^{(*)}$, the coupling constants between $\Upsilon(6S)$ and $B_{(s)}^{(*)}\bar{B}_{(s)}^{(*)}$ can be evaluated by the partial decay widths of $\Upsilon(6S) \to B_{(s)}^{(*)}\bar{B}_{(s)}^{(*)}$. In Table \ref{tab:cc-6S-HH} we list the relevant partial decay widths given in Ref. \cite{Godfrey:2015dia} as well as the corresponding extracted coupling constants.
\begin{table}[htpb]
\centering \caption{The coupling constants of $\Upsilon(6S)$ interacting with $B_{(s)}^{(\ast)} \bar{B}_{(s)}^{(\ast)}$. Here, we also list the corresponding partial decay widths provided in Ref. \cite{Godfrey:2015dia}.\label{tab:cc-6S-HH}}
\begin{tabular}{ccccc}
\toprule[1pt]
Final state  &~& Decay width (MeV) &~& Coupling constant\\
\midrule[0.6pt] %
$B \bar{B}$ &~& 1.32 &~& 0.654\\
$B \bar{B}^{*}$ &~& 7.59 &~& $0.077~\mathrm{GeV}^{-1}$\\
$B^* \bar{B}^*$ &~& 5.89 &~& 0.611\\
$B_s \bar{B}_s$ &~& $1.31\times10^{-3}$ &~& 0.043\\
$B_s \bar{B}_s^{*}$ &~& 0.136 &~& $0.023~\mathrm{GeV}^{-1}$\\
$B_s^* \bar{B}_s^*$ &~& 0.310 &~& 0.354\\
\bottomrule[1pt]
\end{tabular}
\end{table}

The coupling constants relevant to the interactions between $\chi_{bJ}$ and $B_{(s)}^{(*)}\bar{B}_{(s)}^{(*)}$ in the heavy quark limit are related to one gauge coupling $g_1$ given in Eq. (\ref{eqs:SPHH}), i.e.,
\begin{eqnarray}
g_{\chi_{b0} \mathcal{BB}}&=&2\sqrt{3} g_1\sqrt{m_{\chi_{b0}}} m_{\mathcal{B}}, \ \
g_{\chi_{b0} \mathcal{B}^\ast \mathcal{B}^\ast } =\frac{2}{\sqrt{3}}
g_1 \sqrt{m_{\chi_{b0}}} m_{\mathcal{B}^\ast},\nonumber\\
g_{\chi_{b1} \mathcal{B}\mathcal{B}^\ast} &=& 2\sqrt{2} g_1
\sqrt{m_{\chi_{b1}} m_{\mathcal{B}}m_{\mathcal{B}^\ast}}, \ \
g_{\chi_{b2} \mathcal{BB}} =2g_1 \frac{\sqrt{m_{\chi_{b0}}}
}{m_{\mathcal{B}}}, \nonumber\\
g_{\chi_{b2} \mathcal{B} \mathcal{B}^\ast} &=& g_1
\sqrt{\frac{m_{\chi_{b2}}}{m_{\mathcal{B}^\ast}^3 m_{\mathcal{B}}}},
\ \ g_{\chi_{b2} \mathcal{B}^\ast \mathcal{B}^\ast}
=4g_1\sqrt{m_{\chi_{b2}}} m_{\mathcal{B}^\ast},\nonumber
\end{eqnarray}
where $g_1=-\sqrt{m_{\chi_{b0}}\over3}\frac{1}{f_{\chi_{b0}}}$ \cite{Colangelo:2002mj} and $f_{\chi_{b0}}=175 \pm 55$ MeV is the decay constant of $\chi_{b0}$ \cite{Veliev:2010gb}.

Similarly, the coupling constants between $\phi$ or $\omega$ and $B_{(s)}^{(*)}\bar{B}_{(s)}^{(*)}$ can be extracted from Eq. (\ref{eqs:HHV}),
\begin{eqnarray}
g_{B_s B_s \phi} &=&g_{B_s ^\ast
B_s ^\ast \phi}=\frac{\beta g_V}{\sqrt{2}},\nonumber\\
f_{B_s B_s^\ast \phi}
&=&\frac{f_{B_s^\ast B_s^\ast
\phi}}{m_{B_s^\ast}} =\frac{\lambda g_V}{\sqrt{2}},\nonumber\\
g_{B B \omega} &=&g_{B ^\ast
B ^\ast \omega}=\frac{\beta g_V}{2},\nonumber\\
f_{B B^\ast \omega}
&=&\frac{f_{B^\ast B^\ast
\omega}}{m_{B^\ast}} =\frac{\lambda g_V}{2},\nonumber
\end{eqnarray}
with $\beta=0.9$ and $\lambda=0.56 \ \mathrm{GeV}^{-1}$. Additionally, we have $g_V=m_\rho/f_\pi$ along with the pion decay constant $f_\pi=132$ MeV \cite{Cheng:1992xi, Yan:1992gz, Wise:1992hn,Burdman:1992gh}.

With the above preparation, we can evaluate the branching ratios of the $\Upsilon(6S) \to \chi_{bJ} \phi$ and $\Upsilon(6S) \to \chi_{bJ} \omega$ transitions. However, in our model, there still exists a free parameter $\alpha_\Lambda$, which is introduced to parameterize the cutoff $\Lambda$. Since the cutoff $\Lambda$ should not be too far from the physical mass of the exchanged mesons \cite{Cheng:2004ru}, in this work we set the range $0.65 \leq \alpha_\Lambda \leq 1.15$ for $\Upsilon(6S) \to \chi_{bJ} \phi$ transitions and set $0.45 \leq \alpha_\Lambda \leq 1.15$ for $\Upsilon(6S) \to \chi_{bJ} \omega$ transitions to present the numerical results.

\begin{center}
\begin{figure}[htbp]
\scalebox{0.33}{\includegraphics{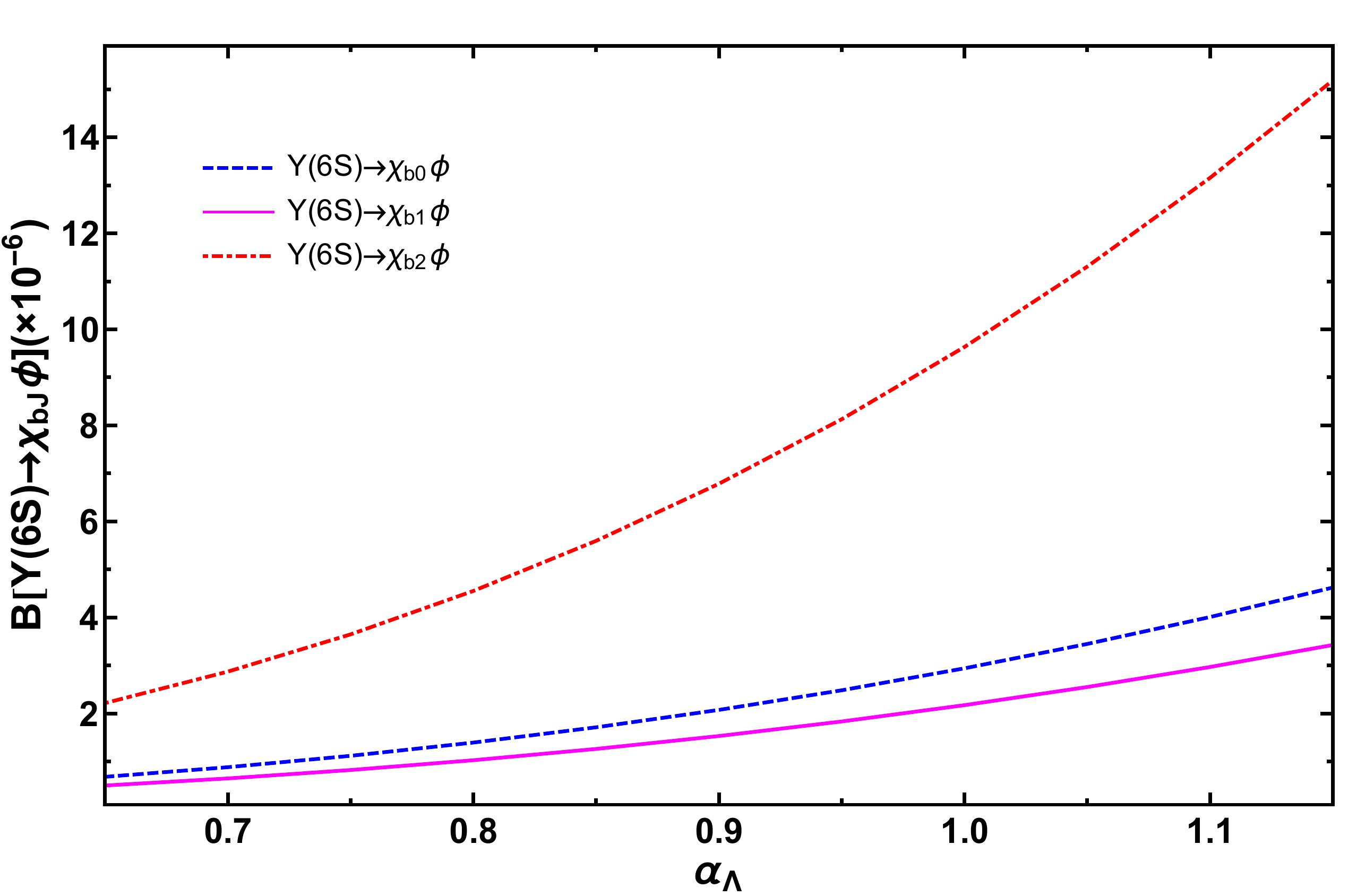}}
\caption{(Color online). The $\alpha_\Lambda$ dependence of the branching ratios of $\Upsilon(6S) \to \chi_{bJ} \phi$.}
\label{fig:B012-phi}
\end{figure}
\end{center}

\begin{center}
\begin{figure}[htbp]
\scalebox{0.33}{\includegraphics{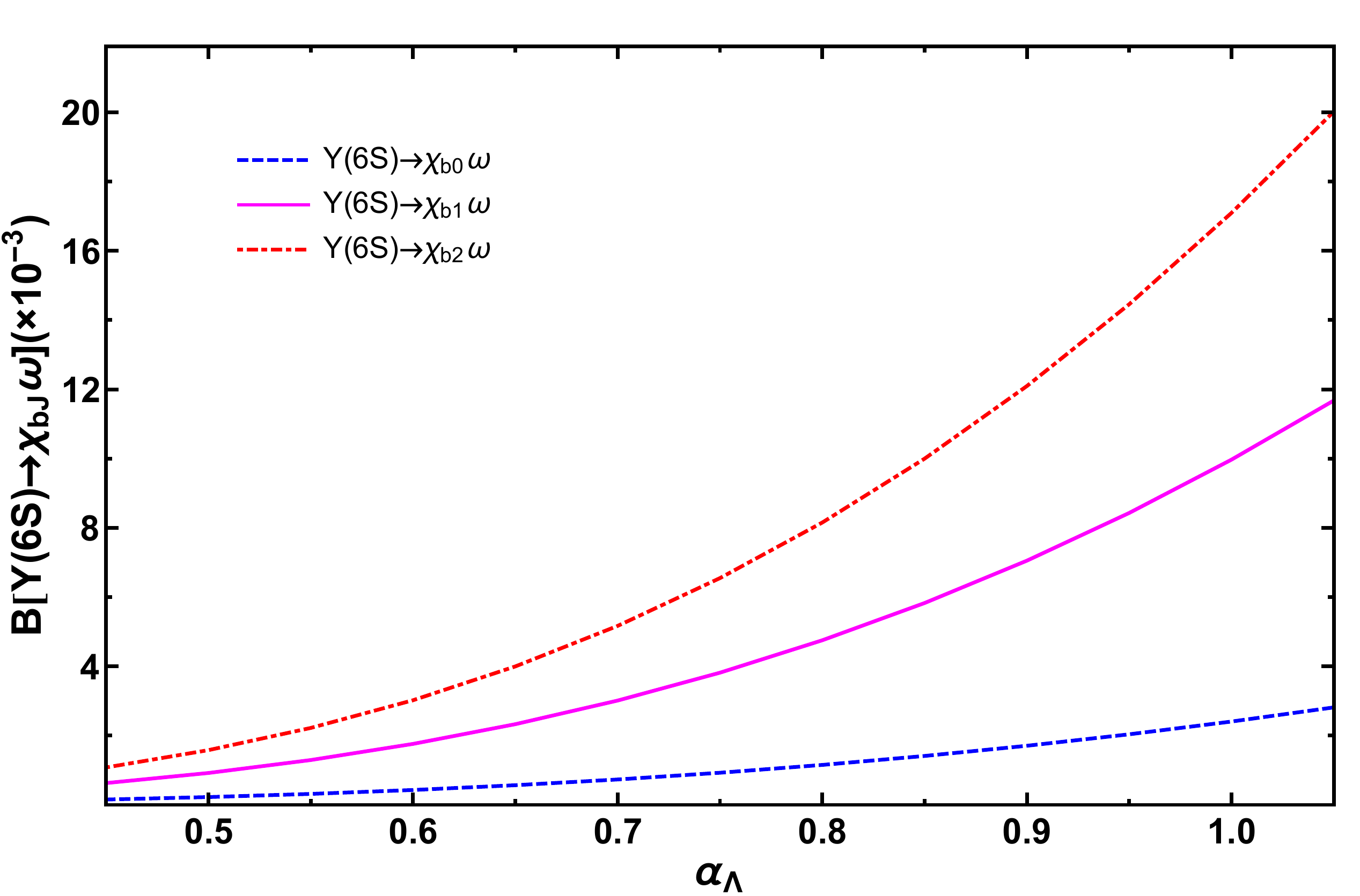}}
\caption{(Color online). The $\alpha_\Lambda$ dependence of the branching ratios of $\Upsilon(6S) \to \chi_{bJ} \omega$.}
\label{fig:B012-omega}
\end{figure}
\end{center}

\begin{center}
\begin{figure}[htbp]
\scalebox{0.33}{\includegraphics{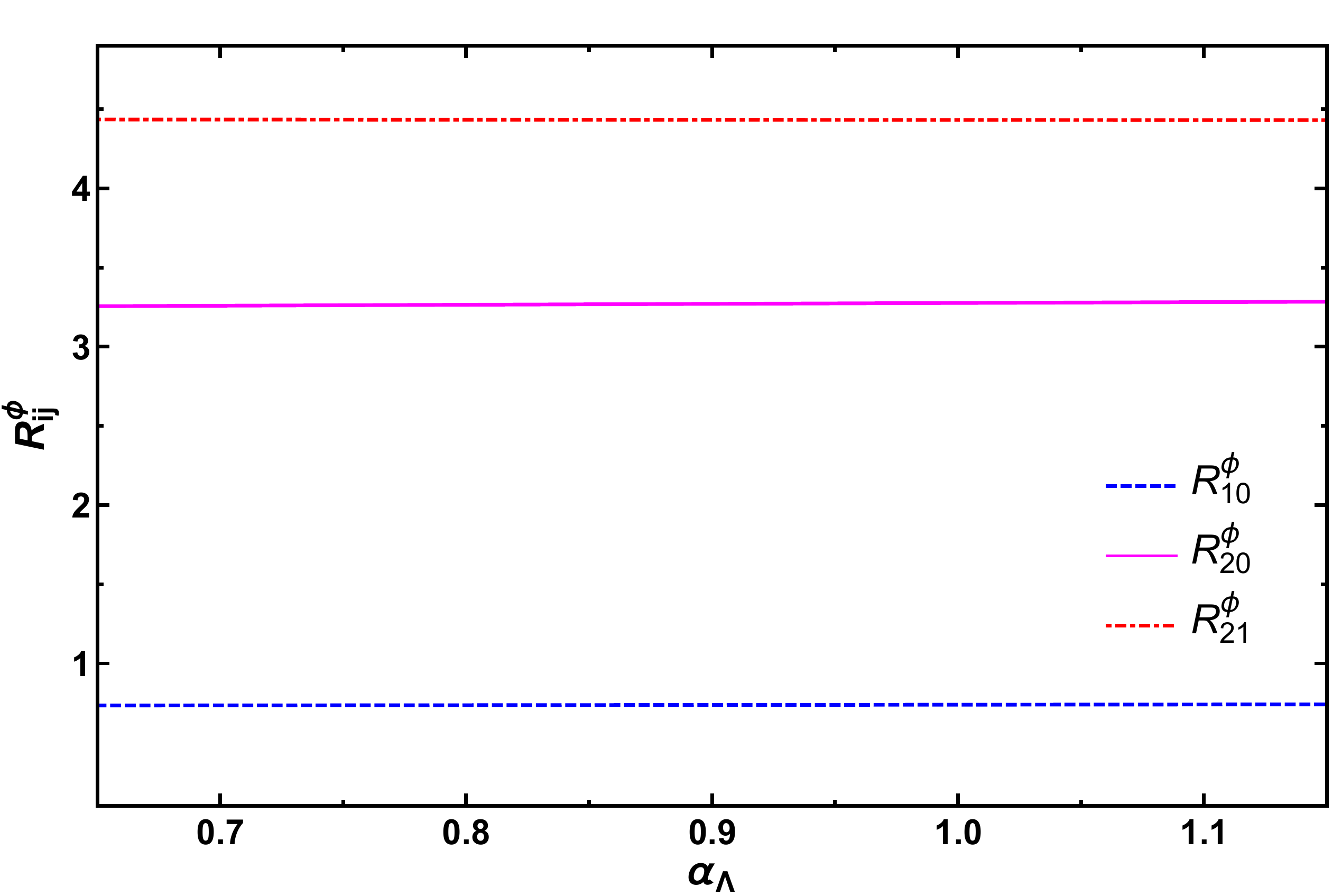}}
\caption{(Color online). The $\alpha_\Lambda$ dependence of the ratios $\mathcal{R}^\phi_{10} = \mathcal{B}[\Upsilon(6S) \to \chi_{b1} \phi] / \mathcal{B}[\Upsilon(6S) \to \chi_{b0} \phi]$, $\mathcal{R}^\phi_{20} = \mathcal{B}[\Upsilon(6S) \to \chi_{b2} \phi] / \mathcal{B}[\Upsilon(6S) \to \chi_{b0} \phi]$ and $\mathcal{R}^\phi_{21} = \mathcal{B}[\Upsilon(6S) \to \chi_{b2} \phi] / \mathcal{B}[\Upsilon(6S) \to \chi_{b1} \phi]$.}
\label{fig:B210-phi}
\end{figure}
\end{center}

\begin{center}
\begin{figure}[htbp]
\scalebox{0.33}{\includegraphics{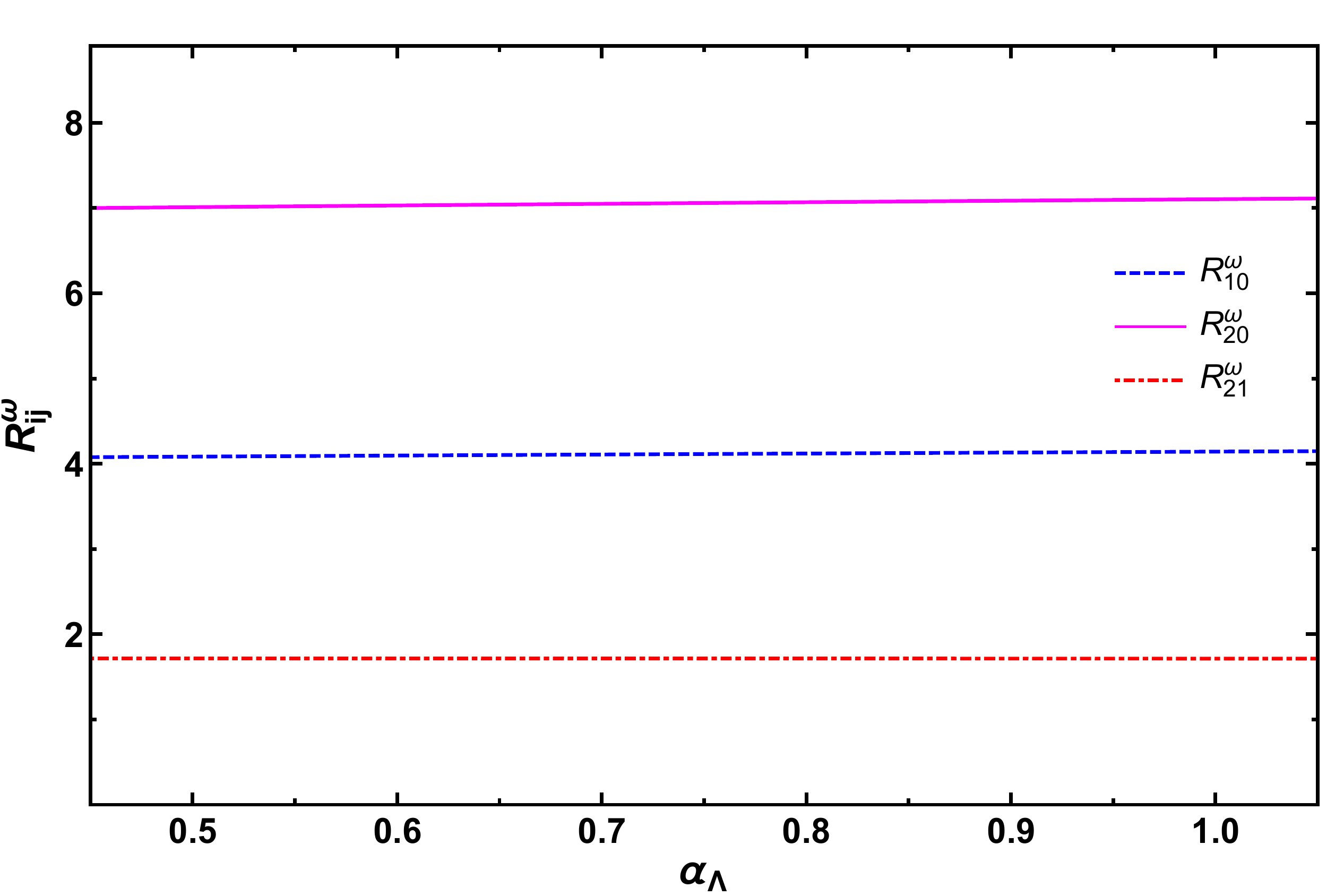}}
\caption{(Color online). The $\alpha_\Lambda$ dependence of the ratios $\mathcal{R}^\omega_{10} = \mathcal{B}[\Upsilon(6S) \to \chi_{b1} \omega] / \mathcal{B}[\Upsilon(6S) \to \chi_{b0} \omega]$, $\mathcal{R}^\omega_{20} = \mathcal{B}[\Upsilon(6S) \to \chi_{b2} \omega] / \mathcal{B}[\Upsilon(6S) \to \chi_{b0} \omega]$ and $\mathcal{R}^\omega_{21} = \mathcal{B}[\Upsilon(6S) \to \chi_{b2} \omega] / \mathcal{B}[\Upsilon(6S) \to \chi_{b1} \omega]$.}
\label{fig:B210-omega}
\end{figure}
\end{center}

In Figs. \ref{fig:B012-phi} and \ref{fig:B012-omega}, we illustrate the $\alpha_\Lambda$ dependence of the branching ratios of $\Upsilon(6S) \to \chi_{bJ} \phi$ and $\Upsilon(6S) \to \chi_{bJ} \omega$, respectively, and in Figs. \ref{fig:B210-phi} and \ref{fig:B210-omega} we present the $\alpha_\Lambda$ dependence of the relative magnitudes among the branching widths of $\Upsilon(6S) \to \chi_{bJ} \phi$ and $\Upsilon(6S) \to \chi_{bJ} \omega$, respectively.

Varying $\alpha_\Lambda$ between 0.65 and 1.15 in $\Upsilon(6S) \to \chi_{bJ} \phi$, we have from Fig. \ref{fig:B012-phi},
\begin{eqnarray}
\mathcal{B}[\Upsilon(6S) \to \chi_{b0} \phi] &=& (0.68 \sim 4.62) \times 10^{-6} , \nonumber\\
\mathcal{B}[\Upsilon(6S) \to \chi_{b1} \phi] &=& (0.50 \sim 3.43) \times 10^{-6} , \nonumber\\
\mathcal{B}[\Upsilon(6S) \to \chi_{b2} \phi] &=& (2.22 \sim 15.18) \times 10^{-6} , \nonumber
\end{eqnarray}
and for $\alpha_\Lambda$ varying from 0.45 to 1.15 in $\Upsilon(6S) \to \chi_{bJ} \omega$, we have from Fig. \ref{fig:B012-omega}
\begin{eqnarray}
\mathcal{B}[\Upsilon(6S) \to \chi_{b0} \omega] &=& (0.15 \sim 2.81) \times 10^{-3} , \nonumber\\
\mathcal{B}[\Upsilon(6S) \to \chi_{b1} \omega] &=& (0.63 \sim 11.68) \times 10^{-3} , \nonumber\\
\mathcal{B}[\Upsilon(6S) \to \chi_{b2} \omega] &=& (1.08 \sim 20.02) \times 10^{-3} . \nonumber
\end{eqnarray}

In addition, some typical values for the relative magnitudes of the predicted branching ratios are obtained from Figs. \ref{fig:B210-phi} and \ref{fig:B210-omega}, which are weakly dependent on the free parameter $\alpha_\Lambda$, i.e.,
\begin{eqnarray}
\mathcal{R}^\phi_{10} &=& \frac{\mathcal{B}[\Upsilon(6S) \to \chi_{b1} \phi]}{\mathcal{B}[\Upsilon(6S) \to \chi_{b0} \phi]} \approx 0.74, \nonumber\\
\mathcal{R}^\phi_{20} &=& \frac{\mathcal{B}[\Upsilon(6S) \to \chi_{b2} \phi]}{\mathcal{B}[\Upsilon(6S) \to \chi_{b0} \phi]} \approx 3.28, \nonumber\\
\mathcal{R}^\phi_{21} &=& \frac{\mathcal{B}[\Upsilon(6S) \to \chi_{b2} \phi]}{\mathcal{B}[\Upsilon(6S) \to \chi_{b1} \phi]} \approx 4.43, \nonumber\\
\mathcal{R}^\omega_{10} &=& \frac{\mathcal{B}[\Upsilon(6S) \to \chi_{b1} \omega]}{\mathcal{B}[\Upsilon(6S) \to \chi_{b0} \omega]} \approx 4.11, \nonumber\\
\mathcal{R}^\omega_{20} &=& \frac{\mathcal{B}[\Upsilon(6S) \to \chi_{b2} \omega]}{\mathcal{B}[\Upsilon(6S) \to \chi_{b0} \omega]} \approx 7.06, \nonumber\\
\mathcal{R}^\omega_{21} &=& \frac{\mathcal{B}[\Upsilon(6S) \to \chi_{b2} \omega]}{\mathcal{B}[\Upsilon(6S) \to \chi_{b1} \omega]} \approx 1.72. \nonumber
\end{eqnarray}

As shown in numerical results on the $\Upsilon(6S) \to \chi_{bJ} \phi$ decays, the partial decay widths of $\Upsilon(6S) \to \chi_{b0} \phi$ and $\Upsilon(6S) \to \chi_{b1} \phi$ are the same order of magnitude, while the partial decay width of $\Upsilon(6S) \to \chi_{b2} \phi$ is one order of magnitude larger than those of $\Upsilon(6S) \to \chi_{b0} \phi$ and $\Upsilon(6S) \to \chi_{b1} \phi$. On the other hand for the $\Upsilon(6S) \to \chi_{bJ} \omega$ decays, the partial decay widths of $\Upsilon(6S) \to \chi_{b1} \omega$ and $\Upsilon(6S) \to \chi_{b2} \omega$ are nearly the same order of magnitude, while the partial decay width of $\Upsilon(6S) \to \chi_{b0} \phi$ is one order of magnitude smaller than those of $\Upsilon(6S) \to \chi_{b1} \omega$ and $\Upsilon(6S) \to \chi_{b2} \omega$.

\section{Summary}


In the past years, the anomalous hadronic transitions like $\Upsilon(5S)\to \Upsilon(nS)\pi^+\pi^-$ ($n=1,2,3$) \cite{Abe:2007tk} and $\Upsilon(5S)\to \chi_{bJ}\omega$ ($J=0,1,2$) \cite{He:2014sqj} were reported by Belle, which has stimulated theorists' interest in revealing the underlying mechanism behind these phenomena  \cite{Chen:2011zv,Meng:2007tk,Meng:2008dd,Simonov:2008qy,Chen:2011qx,Chen:2014ccr,Chen:2011pv}.
As a popular and accepted opinion, the hadronic loop mechanism has been applied to explain why there exist anomalous transitions for $\Upsilon(5S)$  \cite{Chen:2011zv,Meng:2007tk,Meng:2008dd,Simonov:2008qy,Chen:2011qx,Chen:2014ccr,Chen:2011pv}. In addition, more predictions relevant to the $\Upsilon(5S)$ transition were given in Refs. \cite{Meng:2008bq,Wang:2016qmz}.

The main reason to introduce the hadronic loop mechanism is that $\Upsilon(5S)$ is the second observed bottomonium above the $B\bar{B}$ threshold, where the coupled-channel effect may become important, which was tested by the studies in Refs. \cite{Chen:2011zv,Meng:2007tk,Meng:2008dd,Simonov:2008qy,Chen:2011qx,Chen:2014ccr,Chen:2011pv,Meng:2008bq,Wang:2016qmz}.
It is obvious that this is not the end of the whole story.

If the hadronic loop mechanism is a universal mechanism existing in higher bottomonium transitions, we have a reason to believe that this mechanism also plays an important role in higher bottomonium transitions. Considering the similarity between $\Upsilon(6S)$ and $\Upsilon(5S)$, where $\Upsilon(6S)$ is the third bottomomium with open-bottom channels, we have focused on $\Upsilon(6S)\to\chi_{bJ}\phi$ and $\Upsilon(6S)\to\chi_{bJ}\omega$ hadronic decays. Using the hadronic loop mechanism, we have estimated the branching ratios of $\Upsilon(6S)\to\chi_{bJ}\phi$ and $\Upsilon(6S)\to\chi_{bJ}\omega$, which can reach up to $10^{-6}$ and $10^{-3}$, respectively. In the near future, BelleII will be running near the energy range of $\Upsilon(6S)$, which makes BelleII have a great opportunity to find the $\chi_{bJ}\phi$ and $\chi_{bJ}\omega$ decay modes of $\Upsilon(6S)$. If these rare decays are observed, the hadronic loop effects can be further tested.

In this work, we have especially obtained six almost stable ratios $\mathcal{R}^\phi_{10}$, $\mathcal{R}^\phi_{20}$ and $\mathcal{R}^\phi_{21}$ in addition to $\mathcal{R}^\omega_{10}$, $\mathcal{R}^\omega_{20}$ and $\mathcal{R}^\omega_{21}$ reflecting the relative magnitudes of the $\Upsilon(6S)\to\chi_{bJ}\phi$ and $\Upsilon(6S)\to\chi_{bJ}\omega$ decays, which are weakly dependent on our model parameter $\alpha_\Lambda$. Thus, these obtained ratios are important observable quantities. We have also suggested their experimental measurement, which is also a crucial test of our model.

We notice the recent talk of the status of SuperKEKB and the future plan of taking data at the BelleII experiment \cite{Belleii}. Since the collision data on $\Upsilon(6S)$ will be taken, we need to explore the possible interesting research issues about $\Upsilon(6S)$. Our present work is only one step toward the long march.

\section*{Acknowledgments}

This project is supported by the National Natural Science Foundation of China under Grants No.~11222547, No.~11175073, No.~11375240, and No.~11035006, and by Chinese Academy of Sciences under the funding Y104160YQ0 and the agreement No.~2015-BH-02. XL is also supported by the National Program for Support of Young Top-notch Professionals.


\section*{Appendix}
As for the $\Upsilon(6S) \to \chi_{b1} \phi$ transition, the amplitudes corresponding to Fig. \ref{fig:6S-phi-chib1} are
\begin{eqnarray}
\mathcal{M}_{(1-1)} &=& \int \frac{d^4q}{(2\pi)^4} [-i g_{\Upsilon B_s B_s} \epsilon_\Upsilon^\mu ((i k_1)_\mu - (i k_2)_\mu)]\nonumber\\
&&\times [-2 f_{B_s B_s^* \phi} \varepsilon_{\lambda\rho\delta\sigma} \epsilon_\phi^{*\lambda} (i p_2)^\rho ((-i k_1)^\delta - (i q)^\delta)]\nonumber\\
&&\times [i g_{B_s B_s^* \chi_{b1}} \epsilon_{\chi_{b1}}^{*\zeta}] \frac{1}{k_1^2-m_{B_s}^2} \frac{1}{k_2^2-m_{B_s}^2}\nonumber\\
&&\times \frac{-g_\zeta^\sigma + q_\zeta q^\sigma / m_{B_s^*}^2}{q^2-m_{B_s^*}^2}\mathcal{F}^2(q^2),
\end{eqnarray}

\begin{eqnarray}
\mathcal{M}_{(1-2)} &=& \int \frac{d^4q}{(2\pi)^4} [-g_{\Upsilon B_s^* B_s} \varepsilon^{\mu\nu\alpha\beta} \epsilon_{\Upsilon\mu} (-i p_1)_\nu ((i k_2)_\beta - (i k_1)_\beta)]\nonumber\\
&&\times [\epsilon^*_{\phi\lambda} (i g_{B_s^* B_s^* \phi} g^{\delta\sigma} ((-i k_1)^\lambda -(i q)^\lambda)\nonumber\\
&&+ 4 i f_{B_s^* B_s^* \phi} (i p_2)_\rho (g^{\lambda\delta} g^{\rho\sigma} - g^{\lambda\sigma} g^{\rho\delta}))]\nonumber\\
&&\times [i g_{B_s B_s^* \chi_{b1}} \epsilon_{\chi_{b1}}^{*\zeta}] \frac{-g_{\alpha\delta} + k_{1\alpha} k_{1\delta} / m_{B_s^*}^2}{k_1^2-m_{B_s^*}^2}\nonumber\\
&&\times \frac{1}{k_2^2-m_{B_s}^2} \frac{-g_{\sigma\zeta} + q_\sigma q_\zeta / m_{B_s^*}^2}{q^2-m_{B_s^*}^2}\mathcal{F}^2(q^2),
\end{eqnarray}

\begin{eqnarray}
\mathcal{M}_{(1-3)} &=& \int \frac{d^4q}{(2\pi)^4} [g_{\Upsilon B_s B_s^*} \varepsilon^{\mu\nu\alpha\beta} \epsilon_{\Upsilon\mu} (-i p_1)_\nu ((i k_2)_\beta - (i k_1)_\beta)]\nonumber\\
&&\times [-i g_{B_s B_s \phi} \epsilon^*_{\phi\lambda} ((-i k_1)^\lambda -(i q)^\lambda)]\nonumber\\
&&\times [-i g_{B_s^* B_s \chi_{b1}} \epsilon_{\chi_{b1}}^{*\zeta}] \frac{1}{k_1^2-m_{B_s}^2}\nonumber\\
&&\times \frac{-g_{\alpha\zeta} + k_{2\alpha} k_{2\zeta} / m_{B_s^*}^2}{k_2^2-m_{B_s^*}^2} \frac{1}{q^2-m_{B_s}^2}\mathcal{F}^2(q^2),
\end{eqnarray}

\begin{eqnarray}
\mathcal{M}_{(1-4)} &=& \int \frac{d^4q}{(2\pi)^4} [i g_{\Upsilon B_s^* B_s^*} \epsilon_\Upsilon^\mu (g_{\nu\alpha} g_{\mu\beta} (i k_2)^\nu - g_{\mu\alpha} g_{\nu\beta} (i k_1)^\nu\nonumber\\
&& - g_{\alpha\beta}((i k_2)_\mu - (i k_1)_\mu))]\nonumber\\
&&\times [2 f_{B_s^* B_s \phi} \varepsilon_{\lambda\rho\delta\sigma} \epsilon_\phi^{*\lambda} (i p_2)^\rho ((-i k_1)^\delta - (i q)^\delta)]\nonumber\\
&&\times [-i g_{B_s^* B_s \chi_{b1}} \epsilon_{\chi_{b1}}^{*\zeta}] \frac{-g^{\alpha\sigma} + k_1^\alpha k_1^\sigma / m_{B_s^*}^2}{k_1^2-m_{B_s^*}^2}\nonumber\\
&&\times \frac{-g_\zeta^\beta + k_{2\zeta} k_2^\beta / m_{B_s^*}^2}{k_2^2-m_{B_s^*}^2} \frac{1}{q^2-m_{B_s}^2}\mathcal{F}^2(q^2).
\end{eqnarray}

As for the $\Upsilon(6S) \to \chi_{b2} \phi$ transition, the amplitudes corresponding to Fig. \ref{fig:6S-phi-chib2} are
\begin{eqnarray}
\mathcal{M}_{(2-1)} &=& \int \frac{d^4q}{(2\pi)^4} [-i g_{\Upsilon B_s B_s} \epsilon_\Upsilon^\mu ((i k_1)_\mu - (i k_2)_\mu)]\nonumber\\
&&\times [-i g_{B_s B_s \phi} \epsilon^*_{\phi\lambda} ((-i k_1)^\lambda -(i q)^\lambda)]\nonumber\\
&&\times [-g_{B_s B_s \chi_{b2}} \epsilon_{\chi_{b2}}^{*\zeta\eta} (-i q)_\zeta (-i k_2)_\eta]\nonumber\\
&&\times \frac{1}{k_1^2-m_{B_s}^2} \frac{1}{k_2^2-m_{B_s}^2} \frac{1}{q^2-m_{B_s}^2}\mathcal{F}^2(q^2),
\end{eqnarray}

\begin{eqnarray}
\mathcal{M}_{(2-2)} &=& \int \frac{d^4q}{(2\pi)^4} [-i g_{\Upsilon B_s B_s} \epsilon_\Upsilon^\mu ((i k_1)_\mu - (i k_2)_\mu)]\nonumber\\
&&\times [-2 f_{B_s B_s^* \phi} \varepsilon_{\lambda\rho\delta\sigma} \epsilon^{*\lambda}_\phi (i p_2)^\rho ((-i k_1)^\delta - (i q)^\delta)]\nonumber\\
&&\times [-i g_{B_s B_s^* \chi_{b2}} \varepsilon_{\zeta\omega\kappa\xi} \epsilon_{\chi_{b2}}^{*\zeta\eta} (i p_3)^\kappa (-i q)_\eta (-i k_2)^\xi]\nonumber\\
&&\times \frac{1}{k_1^2-m_{B_s}^2} \frac{1}{k_2^2-m_{B_s}^2}\nonumber\\
&&\times \frac{-g^{\sigma\omega} + q^\sigma q^\omega / m_{B_s^*}^2}{q^2-m_{B_s^*}^2}\mathcal{F}^2(q^2),
\end{eqnarray}

\begin{eqnarray}
\mathcal{M}_{(2-3)} &=& \int \frac{d^4q}{(2\pi)^4} [-g_{\Upsilon B_s^* B_s} \varepsilon^{\mu\nu\alpha\beta} \epsilon_{\Upsilon\mu} (-i p_1)_\nu ((i k_2)_\beta - (i k_1)_\beta)]\nonumber\\
&&\times [2 f_{B_s^* B_s \phi} \varepsilon_{\lambda\rho\delta\sigma} \epsilon^{*\lambda}_\phi (i p_2)^\rho ((-i k_1)^\delta - (i q)^\delta)]\nonumber\\
&&\times [-g_{B_s B_s \chi_{b2}} \epsilon_{\chi_{b2}}^{*\zeta\eta} (-i q)_\zeta (-i k_2)_\eta]\nonumber\\
&&\times \frac{-g_\alpha^\sigma + k_{1\alpha} k_1^\sigma / m_{B_s^*}^2}{k_1^2-m_{B_s^*}^2} \frac{1}{k_2^2-m_{B_s}^2} \frac{1}{q^2-m_{B_s}^2}\mathcal{F}^2(q^2),
\end{eqnarray}

\begin{eqnarray}
\mathcal{M}_{(2-4)} &=& \int \frac{d^4q}{(2\pi)^4} [-g_{\Upsilon B_s^* B_s} \varepsilon^{\mu\nu\alpha\beta} \epsilon_{\Upsilon\mu} (-i p_1)_\nu ((i k_2)_\beta - (i k_1)_\beta)]\nonumber\\
&&\times [\epsilon^*_{\phi\lambda} (i g_{B_s^* B_s^* \phi} g^{\delta\sigma} ((-i k_1)^\lambda -(i q)^\lambda)\nonumber\\
&&+ 4 i f_{B_s^* B_s^* \phi} (i p_2)_\rho (g^{\lambda\delta} g^{\rho\sigma} - g^{\lambda\sigma} g^{\rho\delta}))]\nonumber\\
&&\times[-i g_{B_s B_s^* \chi_{b2}} \varepsilon_{\zeta\omega\kappa\xi} \epsilon_{\chi_{b2}}^{*\zeta\eta} (i p_3)^\kappa (-i q)_\eta (-i k_2)^\xi]\nonumber\\
&&\times \frac{-g_{\alpha\delta} - k_{1\alpha} k_{1\delta} / m_{B_s^*}^2}{k_1^2-m_{B_s^*}^2} \frac{1}{k_2^2-m_{B_s}^2}\nonumber\\
&&\times \frac{-g_\sigma^\omega + q_\sigma q^\omega / m_{B_s^*}^2}{q^2-m_{B_s^*}^2}\mathcal{F}^2(q^2),
\end{eqnarray}

\begin{eqnarray}
\mathcal{M}_{(2-5)} &=& \int \frac{d^4q}{(2\pi)^4} [g_{\Upsilon B_s B_s^*} \varepsilon^{\mu\nu\alpha\beta} \epsilon_{\Upsilon\mu} (-i p_1)_\nu ((i k_2)_\beta - (i k_1)_\beta)]\nonumber\\
&&\times [-i g_{B_s B_s \phi} \epsilon^*_{\phi\lambda} ((-i k_1)^\lambda -(i q)^\lambda)]\nonumber\\
&&\times [i g_{B_s^* B_s \chi_{b2}} \varepsilon_{\zeta\omega\kappa\xi} \epsilon_{\chi_{b2}}^{*\zeta\eta} (i p_3)^\kappa (-i q)^\xi (-i k_2)_\eta]\nonumber\\
&&\times \frac{1}{k_1^2-m_{B_s}^2} \frac{-g_\alpha^\omega + k_{2\alpha} k_2^\omega / m_{B_s^*}^2}{k_2^2-m_{B_s^*}^2}\nonumber\\
&&\times \frac{1}{q^2-m_{B_s}^2}\mathcal{F}^2(q^2),
\end{eqnarray}

\begin{eqnarray}
\mathcal{M}_{(2-6)} &=& \int \frac{d^4q}{(2\pi)^4} [g_{\Upsilon B_s B_s^*} \varepsilon^{\mu\nu\alpha\beta} \epsilon_{\Upsilon\mu} (-i p_1)_\nu ((i k_2)_\beta - (i k_1)_\beta)]\nonumber\\
&&\times [-2 f_{B_s B_s^* \phi} \varepsilon_{\lambda\rho\delta\sigma} \epsilon_\phi^{*\lambda} (i p_2)^\rho ((-i k_1)^\delta - (i q)^\delta)]\nonumber\\
&&\times [g_{B_s^* B_s^* \chi_{b2}} \epsilon_{\chi_{b2}}^{*\zeta\eta} (g_{\zeta\kappa} g_{\eta\xi} + g_{\eta\kappa} g_{\zeta\xi})]\nonumber\\
&&\times \frac{1}{k_1^2-m_{B_s}^2} \frac{-g_\alpha^\kappa + k_{2\alpha} k_2^\kappa / m_{B_s^*}^2}{k_2^2-m_{B_s^*}^2}\nonumber\\
&&\times \frac{-g^{\sigma\xi} + q^\sigma q^\xi / m_{B_s^*}^2}{q^2-m_{B_s^*}^2}\mathcal{F}^2(q^2),
\end{eqnarray}

\begin{eqnarray}
\mathcal{M}_{(2-7)} &=& \int \frac{d^4q}{(2\pi)^4} [i g_{\Upsilon B_s^* B_s^*} \epsilon_\Upsilon^\mu (g_{\nu\alpha} g_{\mu\beta} (i k_2)^\nu - g_{\mu\alpha} g_{\nu\beta} (i k_1)^\nu\nonumber\\
&&- g_{\alpha\beta}((i k_2)_\mu - (i k_1)_\mu))]\nonumber\\
&&\times [2 f_{B_s^* B_s \phi} \varepsilon_{\lambda\rho\delta\sigma} \epsilon_\phi^{*\lambda} (i p_2)^\rho ((-i k_1)^\delta - (i q)^\delta)]\nonumber\\
&&\times [i g_{B_s^* B_s \chi_{b2}} \varepsilon_{\zeta\omega\kappa\xi} \epsilon_{\chi_{b2}}^{*\zeta\eta} (i p_3)^\kappa (-i q)^\xi (-i k_2)_\eta]\nonumber]\\
&&\times \frac{-g^{\alpha\sigma} + k_1^\alpha k_1^\sigma / m_{B_s^*}^2}{k_1^2-m_{B_s^*}^2} \frac{-g^{\beta\omega} + k_2^\beta k_2^\omega / m_{B_s^*}^2}{k_2^2-m_{B_s^*}^2}\nonumber\\
&&\times \frac{1}{q^2-m_{B_s}^2}\mathcal{F}^2(q^2),
\end{eqnarray}

\begin{eqnarray}
\mathcal{M}_{(2-8)} &=& \int \frac{d^4q}{(2\pi)^4} [i g_{\Upsilon B_s^* B_s^*} \epsilon_\Upsilon^\mu (g_{\nu\alpha} g_{\mu\beta} (i k_2)^\nu - g_{\mu\alpha} g_{\nu\beta} (i k_1)^\nu\nonumber\\
&&- g_{\alpha\beta}((i k_2)_\mu - (i k_1)_\mu))]\nonumber\\
&&\times [\epsilon^*_{\phi\lambda} (i g_{B_s^* B_s^* \phi} g^{\delta\sigma} ((-i k_1)^\lambda -(i q)^\lambda)\nonumber\\
&&+ 4 i f_{B_s^* B_s^* \phi} (i p_2)_\rho (g^{\lambda\delta} g^{\rho\sigma} - g^{\lambda\sigma} g^{\rho\delta}))]\nonumber\\
&&\times [g_{B_s^* B_s^* \chi_{b2}} \epsilon_{\chi_{b2}}^{*\zeta\eta} (g_{\zeta\kappa} g_{\eta\xi} + g_{\eta\kappa} g_{\zeta\xi})]\nonumber\\
&&\times \frac{-g_\delta^\alpha + k_{1\delta} k_1^\alpha / m_{B_s^*}^2}{k_1^2-m_{B_s^*}^2} \frac{-g^{\beta\kappa} + k_2^\beta k_2^\kappa / m_{B_s^*}^2}{k_2^2-m_{B_s^*}^2}\nonumber\\
&&\times \frac{-g_\sigma^\xi + q_\sigma q^\xi / m_{B_s^*}^2}{q^2-m_{B_s^*}^2}\mathcal{F}^2(q^2).
\end{eqnarray}

As for the $\Upsilon(6S) \to \chi_{b0} \omega$ transition, the amplitudes corresponding to Fig. \ref{fig:6S-omega-chib1} are
\begin{eqnarray}
\mathcal{A}_{(0-1)} &=& \int \frac{d^4q}{(2\pi)^4} [-i g_{\Upsilon B B} \epsilon_\Upsilon^\mu ((i k_1)_\mu - (i k_2)_\mu)]\nonumber\\
&&\times [-i g_{B B \omega} \epsilon^*_{\omega\lambda} ((-i k_1)^\lambda -(i q)^\lambda)] [-g_{B B \chi_{b0}}]\nonumber\\
&&\times \frac{1}{k_1^2-m_{B}^2} \frac{1}{k_2^2-m_{B}^2} \frac{1}{q^2-m_{B}^2}\mathcal{F}^2(q^2),
\end{eqnarray}

\begin{eqnarray}
\mathcal{A}_{(0-2)} &=& \int \frac{d^4q}{(2\pi)^4} [-g_{\Upsilon B^* B} \varepsilon^{\mu\nu\alpha\beta} \epsilon_{\Upsilon\mu} (-i p_1)_\nu ((i k_2)_\beta - (i k_1)_\beta)]\nonumber\\
&&\times [2 f_{B^* B \omega} \varepsilon_{\lambda\rho\delta\sigma} \epsilon_\omega^{*\lambda} (i p_2)^\rho ((-i k_1)^\delta - (i q)^\delta)] [-g_{B B \chi_{b0}}]\nonumber\\
&&\times \frac{-g_\alpha^\sigma + k_{1\alpha} k_1^\sigma / m_{B^*}^2}{k_1^2-m_{B^*}^2} \frac{1}{k_2^2-m_{B}^2} \frac{1}{q^2-m_{B}^2}\mathcal{F}^2(q^2),
\end{eqnarray}

\begin{eqnarray}
\mathcal{A}_{(0-3)} &=& \int \frac{d^4q}{(2\pi)^4} [g_{\Upsilon B B^*} \varepsilon^{\mu\nu\alpha\beta} \epsilon_{\Upsilon\mu} (-i p_1)_\nu ((i k_2)_\beta - (i k_1)_\beta)]\nonumber\\
&&\times [-2 f_{B B^* \omega} \varepsilon_{\lambda\rho\delta\sigma} \epsilon_\omega^{*\lambda} (i p_2)^\rho ((-i k_1)^\delta - (i q)^\delta)]\nonumber\\
&&\times [-g_{B^* B^* \chi_{b0}}]\frac{1}{k_1^2-m_{B}^2} \frac{-g_\alpha^\zeta + k_{2\alpha} k_2^\zeta / m_{B^*}^2}{k_2^2-m_{B^*}^2}\nonumber\\
&&\times \frac{-g^\sigma_\zeta + q^\sigma q_\zeta / m_{B^*}^2 }{q^2-m_{B^*}^2}\mathcal{F}^2(q^2),
\end{eqnarray}

\begin{eqnarray}
\mathcal{A}_{(0-4)} &=& \int \frac{d^4q}{(2\pi)^4} [i g_{\Upsilon B^* B^*} \epsilon_\Upsilon^\mu (g_{\nu\alpha} g_{\mu\beta} (i k_2)^\nu - g_{\mu\alpha} g_{\nu\beta} (i k_1)^\nu\nonumber\\
&& - g_{\alpha\beta}((i k_2)_\mu - (i k_1)_\mu))]\nonumber\\
&&\times [\epsilon^*_{\omega\lambda} (i g_{B^* B^* \omega} g^{\delta\sigma} ((-i k_1)^\lambda -(i q)^\lambda)\nonumber\\
&&+ 4 i f_{B^* B^* \omega} (i p_2)_\rho (g^{\lambda\delta} g^{\rho\sigma} - g^{\lambda\sigma} g^{\rho\delta}))] [-g_{B^* B^* \chi_{b0}}]\nonumber\\
&&\times \frac{-g_\delta^\alpha + k_{1\delta} k_1^\alpha / m_{B^*}^2}{k_1^2-m_{B^*}^2} \frac{-g^{\beta\zeta} + k_2^\beta k_2^\zeta / m_{B^*}^2}{k_2^2-m_{B^*}^2}\nonumber\\
&&\times \frac{-g_{\sigma\zeta} + q_\sigma q_\zeta / m_{B^*}^2}{q^2-m_{B^*}^2}\mathcal{F}^2(q^2).
\end{eqnarray}

As for the $\Upsilon(6S) \to \chi_{b1} \omega$ transition, the amplitudes corresponding to Fig. \ref{fig:6S-omega-chib1} are
\begin{eqnarray}
\mathcal{A}_{(1-1)} &=& \int \frac{d^4q}{(2\pi)^4} [-i g_{\Upsilon B B} \epsilon_\Upsilon^\mu ((i k_1)_\mu - (i k_2)_\mu)]\nonumber\\
&&\times [-2 f_{B B^* \omega} \varepsilon_{\lambda\rho\delta\sigma} \epsilon_\omega^{*\lambda} (i p_2)^\rho ((-i k_1)^\delta - (i q)^\delta)]\nonumber\\
&&\times [i g_{B B^* \chi_{b1}} \epsilon_{\chi_{b1}}^{*\zeta}] \frac{1}{k_1^2-m_{B}^2} \frac{1}{k_2^2-m_{B}^2}\nonumber\\
&&\times \frac{-g_\zeta^\sigma + q_\zeta q^\sigma / m_{B^*}^2}{q^2-m_{B^*}^2}\mathcal{F}^2(q^2),
\end{eqnarray}

\begin{eqnarray}
\mathcal{A}_{(1-2)} &=& \int \frac{d^4q}{(2\pi)^4} [-g_{\Upsilon B^* B} \varepsilon^{\mu\nu\alpha\beta} \epsilon_{\Upsilon\mu} (-i p_1)_\nu ((i k_2)_\beta - (i k_1)_\beta)]\nonumber\\
&&\times [\epsilon^*_{\omega\lambda} (i g_{B^* B^* \omega} g^{\delta\sigma} ((-i k_1)^\lambda -(i q)^\lambda)\nonumber\\
&&+ 4 i f_{B^* B^* \omega} (i p_2)_\rho (g^{\lambda\delta} g^{\rho\sigma} - g^{\lambda\sigma} g^{\rho\delta}))]\nonumber\\
&&\times [i g_{B B^* \chi_{b1}} \epsilon_{\chi_{b1}}^{*\zeta}] \frac{-g_{\alpha\delta} + k_{1\alpha} k_{1\delta} / m_{B^*}^2}{k_1^2-m_{B^*}^2}\nonumber\\
&&\times \frac{1}{k_2^2-m_{B}^2} \frac{-g_{\sigma\zeta} + q_\sigma q_\zeta / m_{B^*}^2}{q^2-m_{B^*}^2}\mathcal{F}^2(q^2),
\end{eqnarray}

\begin{eqnarray}
\mathcal{A}_{(1-3)} &=& \int \frac{d^4q}{(2\pi)^4} [g_{\Upsilon B B^*} \varepsilon^{\mu\nu\alpha\beta} \epsilon_{\Upsilon\mu} (-i p_1)_\nu ((i k_2)_\beta - (i k_1)_\beta)]\nonumber\\
&&\times [-i g_{B B \omega} \epsilon^*_{\omega\lambda} ((-i k_1)^\lambda -(i q)^\lambda)]\nonumber\\
&&\times [-i g_{B^* B \chi_{b1}} \epsilon_{\chi_{b1}}^{*\zeta}] \frac{1}{k_1^2-m_{B}^2}\nonumber\\
&&\times \frac{-g_{\alpha\zeta} + k_{2\alpha} k_{2\zeta} / m_{B^*}^2}{k_2^2-m_{B^*}^2} \frac{1}{q^2-m_{B}^2}\mathcal{F}^2(q^2),
\end{eqnarray}

\begin{eqnarray}
\mathcal{A}_{(1-4)} &=& \int \frac{d^4q}{(2\pi)^4} [i g_{\Upsilon B^* B^*} \epsilon_\Upsilon^\mu (g_{\nu\alpha} g_{\mu\beta} (i k_2)^\nu - g_{\mu\alpha} g_{\nu\beta} (i k_1)^\nu\nonumber\\
&& - g_{\alpha\beta}((i k_2)_\mu - (i k_1)_\mu))]\nonumber\\
&&\times [2 f_{B^* B \omega} \varepsilon_{\lambda\rho\delta\sigma} \epsilon_\omega^{*\lambda} (i p_2)^\rho ((-i k_1)^\delta - (i q)^\delta)]\nonumber\\
&&\times [-i g_{B^* B \chi_{b1}} \epsilon_{\chi_{b1}}^{*\zeta}] \frac{-g^{\alpha\sigma} + k_1^\alpha k_1^\sigma / m_{B^*}^2}{k_1^2-m_{B^*}^2}\nonumber\\
&&\times \frac{-g_\zeta^\beta + k_{2\zeta} k_2^\beta / m_{B^*}^2}{k_2^2-m_{B^*}^2} \frac{1}{q^2-m_{B}^2}\mathcal{F}^2(q^2).
\end{eqnarray}

As for the $\Upsilon(6S) \to \chi_{b2} \omega$ transition, the amplitudes corresponding to Fig. \ref{fig:6S-omega-chib2} are
\begin{eqnarray}
\mathcal{A}_{(2-1)} &=& \int \frac{d^4q}{(2\pi)^4} [-i g_{\Upsilon B B} \epsilon_\Upsilon^\mu ((i k_1)_\mu - (i k_2)_\mu)]\nonumber\\
&&\times [-i g_{B B \omega} \epsilon^*_{\omega\lambda} ((-i k_1)^\lambda -(i q)^\lambda)]\nonumber\\
&&\times [-g_{B B \chi_{b2}} \epsilon_{\chi_{b2}}^{*\zeta\eta} (-i q)_\zeta (-i k_2)_\eta]\nonumber\\
&&\times \frac{1}{k_1^2-m_{B}^2} \frac{1}{k_2^2-m_{B}^2} \frac{1}{q^2-m_{B}^2}\mathcal{F}^2(q^2),
\end{eqnarray}

\begin{eqnarray}
\mathcal{A}_{(2-2)} &=& \int \frac{d^4q}{(2\pi)^4} [-i g_{\Upsilon B B} \epsilon_\Upsilon^\mu ((i k_1)_\mu - (i k_2)_\mu)]\nonumber\\
&&\times [-2 f_{B B^* \omega} \varepsilon_{\lambda\rho\delta\sigma} \epsilon^{*\lambda}_\omega (i p_2)^\rho ((-i k_1)^\delta - (i q)^\delta)]\nonumber\\
&&\times [-i g_{B B^* \chi_{b2}} \varepsilon_{\zeta\omega\kappa\xi} \epsilon_{\chi_{b2}}^{*\zeta\eta} (i p_3)^\kappa (-i q)_\eta (-i k_2)^\xi]\nonumber\\
&&\times \frac{1}{k_1^2-m_{B}^2} \frac{1}{k_2^2-m_{B}^2}\nonumber\\
&&\times \frac{-g^{\sigma\omega} + q^\sigma q^\omega / m_{B^*}^2}{q^2-m_{B^*}^2}\mathcal{F}^2(q^2),
\end{eqnarray}

\begin{eqnarray}
\mathcal{A}_{(2-3)} &=& \int \frac{d^4q}{(2\pi)^4} [-g_{\Upsilon B^* B} \varepsilon^{\mu\nu\alpha\beta} \epsilon_{\Upsilon\mu} (-i p_1)_\nu ((i k_2)_\beta - (i k_1)_\beta)]\nonumber\\
&&\times [2 f_{B^* B \omega} \varepsilon_{\lambda\rho\delta\sigma} \epsilon^{*\lambda}_\omega (i p_2)^\rho ((-i k_1)^\delta - (i q)^\delta)]\nonumber\\
&&\times [-g_{B B \chi_{b2}} \epsilon_{\chi_{b2}}^{*\zeta\eta} (-i q)_\zeta (-i k_2)_\eta]\nonumber\\
&&\times \frac{-g_\alpha^\sigma + k_{1\alpha} k_1^\sigma / m_{B^*}^2}{k_1^2-m_{B^*}^2} \frac{1}{k_2^2-m_{B}^2} \frac{1}{q^2-m_{B}^2}\mathcal{F}^2(q^2),
\end{eqnarray}

\begin{eqnarray}
\mathcal{A}_{(2-4)} &=& \int \frac{d^4q}{(2\pi)^4} [-g_{\Upsilon B^* B} \varepsilon^{\mu\nu\alpha\beta} \epsilon_{\Upsilon\mu} (-i p_1)_\nu ((i k_2)_\beta - (i k_1)_\beta)]\nonumber\\
&&\times [\epsilon^*_{\omega\lambda} (i g_{B^* B^* \omega} g^{\delta\sigma} ((-i k_1)^\lambda -(i q)^\lambda)\nonumber\\
&&+ 4 i f_{B^* B^* \omega} (i p_2)_\rho (g^{\lambda\delta} g^{\rho\sigma} - g^{\lambda\sigma} g^{\rho\delta}))]\nonumber\\
&&\times[-i g_{B B^* \chi_{b2}} \varepsilon_{\zeta\omega\kappa\xi} \epsilon_{\chi_{b2}}^{*\zeta\eta} (i p_3)^\kappa (-i q)_\eta (-i k_2)^\xi]\nonumber\\
&&\times \frac{-g_{\alpha\delta} - k_{1\alpha} k_{1\delta} / m_{B^*}^2}{k_1^2-m_{B^*}^2} \frac{1}{k_2^2-m_{B}^2}\nonumber\\
&&\times \frac{-g_\sigma^\omega + q_\sigma q^\omega / m_{B^*}^2}{q^2-m_{B^*}^2}\mathcal{F}^2(q^2),
\end{eqnarray}

\begin{eqnarray}
\mathcal{A}_{(2-5)} &=& \int \frac{d^4q}{(2\pi)^4} [g_{\Upsilon B B^*} \varepsilon^{\mu\nu\alpha\beta} \epsilon_{\Upsilon\mu} (-i p_1)_\nu ((i k_2)_\beta - (i k_1)_\beta)]\nonumber\\
&&\times [-i g_{B B \omega} \epsilon^*_{\omega\lambda} ((-i k_1)^\lambda -(i q)^\lambda)]\nonumber\\
&&\times [i g_{B^* B \chi_{b2}} \varepsilon_{\zeta\omega\kappa\xi} \epsilon_{\chi_{b2}}^{*\zeta\eta} (i p_3)^\kappa (-i q)^\xi (-i k_2)_\eta]\nonumber\\
&&\times \frac{1}{k_1^2-m_{B}^2} \frac{-g_\alpha^\omega + k_{2\alpha} k_2^\omega / m_{B^*}^2}{k_2^2-m_{B^*}^2}\nonumber\\
&&\times \frac{1}{q^2-m_{B}^2}\mathcal{F}^2(q^2),
\end{eqnarray}

\begin{eqnarray}
\mathcal{A}_{(2-6)} &=& \int \frac{d^4q}{(2\pi)^4} [g_{\Upsilon B B^*} \varepsilon^{\mu\nu\alpha\beta} \epsilon_{\Upsilon\mu} (-i p_1)_\nu ((i k_2)_\beta - (i k_1)_\beta)]\nonumber\\
&&\times [-2 f_{B B^* \omega} \varepsilon_{\lambda\rho\delta\sigma} \epsilon_\omega^{*\lambda} (i p_2)^\rho ((-i k_1)^\delta - (i q)^\delta)]\nonumber\\
&&\times [g_{B^* B^* \chi_{b2}} \epsilon_{\chi_{b2}}^{*\zeta\eta} (g_{\zeta\kappa} g_{\eta\xi} + g_{\eta\kappa} g_{\zeta\xi})]\nonumber\\
&&\times \frac{1}{k_1^2-m_{B}^2} \frac{-g_\alpha^\kappa + k_{2\alpha} k_2^\kappa / m_{B^*}^2}{k_2^2-m_{B^*}^2}\nonumber\\
&&\times \frac{-g^{\sigma\xi} + q^\sigma q^\xi / m_{B^*}^2}{q^2-m_{B^*}^2}\mathcal{F}^2(q^2),
\end{eqnarray}

\begin{eqnarray}
\mathcal{A}_{(2-7)} &=& \int \frac{d^4q}{(2\pi)^4} [i g_{\Upsilon B^* B^*} \epsilon_\Upsilon^\mu (g_{\nu\alpha} g_{\mu\beta} (i k_2)^\nu - g_{\mu\alpha} g_{\nu\beta} (i k_1)^\nu\nonumber\\
&&- g_{\alpha\beta}((i k_2)_\mu - (i k_1)_\mu))]\nonumber\\
&&\times [2 f_{B^* B \omega} \varepsilon_{\lambda\rho\delta\sigma} \epsilon_\omega^{*\lambda} (i p_2)^\rho ((-i k_1)^\delta - (i q)^\delta)]\nonumber\\
&&\times [i g_{B^* B \chi_{b2}} \varepsilon_{\zeta\omega\kappa\xi} \epsilon_{\chi_{b2}}^{*\zeta\eta} (i p_3)^\kappa (-i q)^\xi (-i k_2)_\eta]\nonumber]\\
&&\times \frac{-g^{\alpha\sigma} + k_1^\alpha k_1^\sigma / m_{B^*}^2}{k_1^2-m_{B^*}^2} \frac{-g^{\beta\omega} + k_2^\beta k_2^\omega / m_{B^*}^2}{k_2^2-m_{B^*}^2}\nonumber\\
&&\times \frac{1}{q^2-m_{B}^2}\mathcal{F}^2(q^2),
\end{eqnarray}

\begin{eqnarray}
\mathcal{A}_{(2-8)} &=& \int \frac{d^4q}{(2\pi)^4} [i g_{\Upsilon B^* B^*} \epsilon_\Upsilon^\mu (g_{\nu\alpha} g_{\mu\beta} (i k_2)^\nu - g_{\mu\alpha} g_{\nu\beta} (i k_1)^\nu\nonumber\\
&&- g_{\alpha\beta}((i k_2)_\mu - (i k_1)_\mu))]\nonumber\\
&&\times [\epsilon^*_{\omega\lambda} (i g_{B^* B^* \omega} g^{\delta\sigma} ((-i k_1)^\lambda -(i q)^\lambda)\nonumber\\
&&+ 4 i f_{B^* B^* \omega} (i p_2)_\rho (g^{\lambda\delta} g^{\rho\sigma} - g^{\lambda\sigma} g^{\rho\delta}))]\nonumber\\
&&\times [g_{B^* B^* \chi_{b2}} \epsilon_{\chi_{b2}}^{*\zeta\eta} (g_{\zeta\kappa} g_{\eta\xi} + g_{\eta\kappa} g_{\zeta\xi})]\nonumber\\
&&\times \frac{-g_\delta^\alpha + k_{1\delta} k_1^\alpha / m_{B^*}^2}{k_1^2-m_{B^*}^2} \frac{-g^{\beta\kappa} + k_2^\beta k_2^\kappa / m_{B^*}^2}{k_2^2-m_{B^*}^2}\nonumber\\
&&\times \frac{-g_\sigma^\xi + q_\sigma q^\xi / m_{B^*}^2}{q^2-m_{B^*}^2}\mathcal{F}^2(q^2).
\end{eqnarray}

\vfil

\end{document}